\documentclass[prd,superscriptaddress,unsortedaddress,twocolumn,showpacs,floatfixamsmath,amssymb]{revtex4}
 \newcommand{\beq}{\begin{equation}}
 \newcommand{\eeq}{\end{equation}}
 \newcommand{\bqa}{\begin{eqnarray}}
 \newcommand{\eqa}{\end{eqnarray}}

\def\ket#1{\left| #1\right\rangle}

\newcommand{\Kz}{\mbox{$K^{0}$}}
\newcommand{\Kzbar}{\mbox{$\overline{K^{0}}$}}

\newcommand{\KL}{\mbox{$K_{L}$}}
\newcommand{\KS}{\mbox{$K_{S}$}}
\newcommand{\Kl}{\mbox{$K_{L}$}}

\newcommand{\ptsq}{\mbox{$p_T^2$}}

\newcommand{\tauS}{\mbox{$\tau_{S}$}}

\newcommand{\delm}{\mbox{$\Delta m$}}

\def\epe{\epsilon'\!/\epsilon}

\newcommand{\epspr}{\mbox{$\epsilon^\prime$}}

\newcommand{\epsrat}{\mbox{$\epsilon^{\prime}\!/\epsilon$}}
\newcommand{\reepoe}{\mbox{$Re(\epsrat)$}}
\newcommand{\imepoe}{\mbox{$Im(\epsrat)$}}
\newcommand{\ktev}{\mbox{KTeV}}
\newcommand{\etapm}{\mbox{$\eta_{+-}$}}
\newcommand{\etazz}{\mbox{$\eta_{00}$}}
\newcommand{\phipm}{\mbox{$\phi_{+-}$}}
\newcommand{\phizz}{\mbox{$\phi_{00}$}}
\newcommand{\phisw}{\mbox{$\phi_{SW}$}}
\newcommand{\phiep}{\mbox{$\phi_{\epsilon}$}}
\newcommand{\delphi}{\mbox{$\Delta \phi$}}

\def\kchrg{K\to\pi^+\pi^-}
\def\kneut{K\to\pi^0\pi^0}
\def\kppp{K_{L}\to \pi^0\pi^0\pi^0}
\def\kpi0{K_{L}\to 3\pi^0}
\def\ke3{K_{L}\to\pi^{\pm}e^{\mp}\nu}
\def\km3{K_{L}\to\pi^{\pm}\mu^{\mp}\nu}
\def\k2pi{K_{L} \to \pi^+\pi^-}

\newcommand{\Kpp}{\mbox{$K\rightarrow\pi\pi$}}
\newcommand{\Kpm}{\mbox{$K\rightarrow\pi^{+}\pi^{-}$}}
\newcommand{\Kzz}{\mbox{$K\rightarrow\pi^{0}\pi^{0}$}}

\newcommand{\Kethree}{\mbox{$K_{e3}$}}
\newcommand{\Kmuthree}{\mbox{$K_{\mu 3}$}}

\newcommand{\KLpm}{\mbox{$K_{L}\rightarrow\pi^{+}\pi^{-}$}}
\newcommand{\KSpm}{\mbox{$K_{S}\rightarrow\pi^{+}\pi^{-}$}}
\newcommand{\KLzz}{\mbox{$K_{L}\rightarrow\pi^{0}\pi^{0}$}}
\newcommand{\KSzz}{\mbox{$K_{S}\rightarrow\pi^{0}\pi^{0}$}}

\newcommand{\KLpmz}{\mbox{$K_{L}\rightarrow \pi^{+}\pi^{-}\pi^{0}$}}
\newcommand{\Kzzz}{\mbox{$K_L \rightarrow \pi^{0}\pi^{0}\pi^{0}$}}
\newcommand{\Kpmz}{\mbox{$K \rightarrow \pi^{+}\pi^{-}\pi^{0}$}}

\newcommand{\KLmu}{\mbox{$\Kl\rightarrow \pi^{\pm} \mu^{\mp}\nu_\mu$}}

\newcommand{\Lppi}{\mbox{$\Lambda \rightarrow p \pi^-$}}

\newcommand{\piz}{\mbox{$\pi^{0}$}}

\newcommand{\ppc}{\mbox{$\pi^{+}\pi^{-}$}}

\newcommand{\ppn}{\mbox{$\pi^{0}\pi^{0}$}}
\newcommand{\pppm}{\pi^{+}\pi^{-}}
\newcommand{\pzpz}{\pi^{0}\pi^{0}}
\newcommand{\kmth}{\mbox{$\pi^{\pm}\mu^{\mp}\nu$}}
\newcommand{\zzz}{\mbox{$\pi^{0}\pi^{0}\pi^{0}$}}

\newcommand{\chisqzz}{\chi^2_{\pi^0}}
\newcommand{\chisqvtx}{\chi^2_{vtx}}

\newcommand{\chisqshape}{\chi^2_{\gamma}}
\newcommand{\dzvtx}{\Delta z_{vtx}}

\newcommand{\eu}{ \times 10^{-4}}

\newcommand{\upk}{ {\rm GeV}/$c$ }
\newcommand{\umass}{ {\rm MeV}/$c$^2}

\newcommand{\degs}{^{\circ}}
\newcommand{\delmunits}{\mbox{$\times 10^{6}~\hbar {\rm s}^{-1}$}}
\newcommand{\tausunits}{\mbox{$\times 10^{-12}~{\rm s}$}}

\newcommand{\fminus}{\mbox{$|f_{-}(70~{\rm GeV}/c)|$}}
\newcommand{\sigdz}{\mbox{$\sigma_{\Delta z}$}}

\def\dmswval{5269.9}
\def\dmswvalr{5270}
\def\tsswval{89.623}
\def\tsswvalr{89.62}
\def\phswswval{43.419}
\def\dmswerr{  12.3}
\def\dmswerrr{12}
\def\tsswerr{ 0.047}
\def\tsswerrr{0.05}
\def\phswswerr{ 0.058}
\def\dmcptval{5279.7}
\def\tscptval{89.589}
\def\phcptval{ 43.86}
\def\reecptval{ 21.10}
\def\imecptval{-17.20}
\def\dphcptval{ 0.30}
\def\dmcpterr{  19.5}
\def\tscpterr{ 0.070}
\def\phcpterr{  0.63}
\def\reecpterr{  3.43}
\def\imecpterr{ 20.20}
\def\dphcpterr{ 0.35}
\def\dphcpterrsy{ 0.31}
\def\dmtsswcor{ -67.0}
\def\dmtscptcor{ -85.8}
\def\dmphcptcor{  82.8}
\def\dmreecptcor{  -6.6}
\def\dmimecptcor{   2.6}
\def\tsphcptcor{ -76.5}
\def\tsreecptcor{   6.8}
\def\tsimecptcor{  -1.0}
\def\phreecptcor{   2.5}
\def\phimecptcor{  -4.1}
\def\reeimecptcor{ -46.5}
\def\phpmval{ 43.76}
\def\phpmerr{  0.64}
\def\phzzval{ 44.06}
\def\phzzerr{  0.68}
\def\dmcpterrsy{  14.7}
\def\tscpterrsy{ 0.056}
\def\phcpterrsy{  0.49}
\def\reecpterrsy{  3.17}
\def\imecpterrsy{ 18.06}
\def\dphswval{  0.40}
\def\dphswerr{  0.56}
\def\dphswerrsy{  0.42}
\def\phswcptval{43.461}
\def\phswcpterr{ 0.098}
\def\phswcpterrsy{ 0.070}

\def\dmkzkbimv{-1.5} 
\def\dmkzkbime{1.6} 
\def\dmkzkbfrefv{25.1} 
\def\dmkzkbfrefe{22.5} 
\def\dmkzkbimfv{-1.5} 
\def\dmkzkbimfe{1.6} 
\def\dmkzkblimit{4.8} 

\usepackage{epsfig}
\usepackage{graphicx}% Include figure files
\usepackage{dcolumn}% Align table columns on decimal point
\usepackage{bm}% bold math

\begin{document}
%\begin{linenumbers}

\preprint{draft}

\title{Precise Measurements of Direct CP Violation, 
CPT Symmetry, and Other 
Parameters in the Neutral Kaon System}

\newcommand{\UAz}{University of Arizona, Tucson, Arizona 85721}
\newcommand{\UCLA}{University of California at Los Angeles, Los Angeles,
                    California 90095} 
\newcommand{\Campinas}{Universidade Estadual de Campinas, Campinas, 
                       Brazil 13083-970}
\newcommand{\EFI}{The Enrico Fermi Institute, The University of Chicago, 
                  Chicago, Illinois 60637}
\newcommand{\UB}{University of Colorado, Boulder, Colorado 80309}
\newcommand{\ELM}{Elmhurst College, Elmhurst, Illinois 60126}
\newcommand{\FNAL}{Fermi National Accelerator Laboratory, 
                   Batavia, Illinois 60510}
\newcommand{\Osaka}{Osaka University, Toyonaka, Osaka 560-0043 Japan} 
\newcommand{\Rice}{Rice University, Houston, Texas 77005}
\newcommand{\SaoPaolo}{Universidade de S\~ao Paulo, S\~ao Paulo, Brazil 05315-970}
\newcommand{\UVa}{The Department of Physics and Institute of Nuclear and 
                  Particle Physics, University of Virginia, 
                  Charlottesville, Virginia 22901}
\newcommand{\UW}{University of Wisconsin, Madison, Wisconsin 53706}
\newcommand{\DESY}{DESY, Hamburg, Germany}

\affiliation{\UAz}
\affiliation{\UCLA}
\affiliation{\Campinas}
\affiliation{\EFI}
\affiliation{\UB}
\affiliation{\ELM}
\affiliation{\FNAL}
\affiliation{\Osaka}
\affiliation{\Rice}
\affiliation{\SaoPaolo}
\affiliation{\UVa}
\affiliation{\UW}

\author{E.~Abouzaid}	  \affiliation{\EFI}
\author{M.~Arenton}       \affiliation{\UVa}
\author{A.R.~Barker}      \altaffiliation[Deceased.]{ } \affiliation{\UB}
\author{M.~Barrio}        \affiliation{\EFI}
\author{L.~Bellantoni}    \affiliation{\FNAL}
\author{E.~Blucher}       \affiliation{\EFI}
\author{G.J.~Bock}        \affiliation{\FNAL}
\author{C.~Bown}          \affiliation{\EFI}
\author{E.~Cheu}          \affiliation{\UAz}
\author{R.~Coleman}       \affiliation{\FNAL}
\author{M.D.~Corcoran}    \affiliation{\Rice}
\author{B.~Cox}           \affiliation{\UVa}
\author{A.R.~Erwin}       \affiliation{\UW}
\author{C.O.~Escobar}     \affiliation{\Campinas}  %%Consider after Ke4 paper
\author{A.~Glazov}        
   \altaffiliation[Permanent address ]{\DESY}
   \affiliation{\EFI}
\author{A.~Golossanov}    \affiliation{\UVa} %%Remove March 08
\author{R.A.~Gomes}       \affiliation{\Campinas}
\author{P. Gouffon}       \affiliation{\SaoPaolo}
\author{J.~Graham}        \affiliation{\EFI}
\author{J.~Hamm}          \affiliation{\UW}
\author{Y.B.~Hsiung}      \affiliation{\FNAL}
\author{D.A.~Jensen}      \affiliation{\FNAL}
\author{R.~Kessler}       \affiliation{\EFI}
\author{K.~Kotera}	  \affiliation{\Osaka}
\author{J.~LaDue}         \affiliation{\UB}
\author{A.~Ledovskoy}     \affiliation{\UVa}
\author{P.L.~McBride}     \affiliation{\FNAL}

\author{E.~Monnier}
   \altaffiliation[Permanent address ]{C.P.P. Marseille/C.N.R.S., France}
   \affiliation{\EFI}  %% Doug will ping him 

\author{H.~Nguyen}       \affiliation{\FNAL}
\author{R.~Niclasen}     \affiliation{\UB}
\author{D.G.~Phillips~II} \affiliation{\UVa}
\author{V.~Prasad}       \affiliation{\EFI} %%case by case 832 only
\author{X.R.~Qi}         \affiliation{\FNAL} %%June 06
\author{E.J.~Ramberg}    \affiliation{\FNAL}
\author{R.E.~Ray}        \affiliation{\FNAL}
\author{M.~Ronquest}     \affiliation{\UVa}
\author{A.~Roodman}      \affiliation{\EFI}
\author{E.~Santos}       \affiliation{\SaoPaolo}
\author{P.~Shanahan}     \affiliation{\FNAL}
\author{P.S.~Shawhan}    \affiliation{\EFI}
\author{W.~Slater}       \affiliation{\UCLA}
\author{D.~Smith}        \affiliation{\UVa}
\author{N.~Solomey}      \affiliation{\EFI}
\author{E.C.~Swallow}    \affiliation{\EFI}\affiliation{\ELM}
\author{S.A.~Taegar}     \affiliation{\UAz}
\author{P.A.~Toale}      \affiliation{\UB}
\author{R.~Tschirhart}   \affiliation{\FNAL}
\author{Y.W.~Wah}        \affiliation{\EFI}
\author{J.~Wang}         \affiliation{\UAz}
\author{H.B.~White}      \affiliation{\FNAL}
\author{J.~Whitmore}     \affiliation{\FNAL}
\author{M.~J.~Wilking}      \affiliation{\UB}
\author{B.~Winstein}     \affiliation{\EFI}
\author{R.~Winston}      \affiliation{\EFI}
\author{E.T.~Worcester}  \affiliation{\EFI}
\author{T.~Yamanaka}     \affiliation{\Osaka}
\author{E.~D.~Zimmerman} \affiliation{\UB}
\author{R.F.~Zukanovich} \affiliation{\SaoPaolo}

\date{\today}% It is always \today, today,

\begin{abstract}

We present precise tests of CP and CPT symmetry
based on the full dataset of $\Kpp$ decays collected by the KTeV experiment 
at Fermi National Accelerator Laboratory during 1996, 1997, and 1999.
This dataset contains 16 million $\kneut$ and 69 million $\kchrg$ decays.  
We measure the direct CP violation parameter $\reepoe$ = (19.2 $\pm 2.1)\eu$.
We find the $K_L$-$K_S$ mass difference $\delm = (\dmswvalr \pm \dmswerrr)\delmunits$
and the $K_S$ lifetime $\tauS$ = ($\tsswvalr \pm \tsswerrr)\tausunits$.  We also
measure several parameters that test CPT invariance.
We find the difference between the phase of the indirect CP violation parameter, $\epsilon$,
and the superweak phase, $\phiep - \phi_{SW} = (\dphswval \pm \dphswerr)\degs$.
%We find the phase of the indirect CP violation parameter 
%$\epsilon$, $\phiep$~=~(44.09 $\pm$ 1.00)$\degs$.  
We measure the difference of 
the relative phases between the CP violating and CP conserving decay amplitudes for 
$\Kpm$ ($\phipm$) and for $\Kzz$ ($\phizz$), $\delphi = (\dphcptval \pm \dphcpterr)\degs$.
From these phase measurements, we place a limit on the mass difference between
$\Kz$ and $\Kzbar$, $\Delta M <~\dmkzkblimit~\times~10^{-19}~\mbox{GeV/$c^2$ at } 95\%~\mbox{C.L.}$  
These results are consistent with those of other experiments, 
our own earlier measurements, and CPT symmetry.

\end{abstract}

\pacs{11.30.Er, 13.25.Es, 14.40.Df}% PACS, the Physics and Astronomy

\maketitle

\section{\label{sec:intro}Introduction}

\begin{figure}
\begin{center}
\epsfig{file=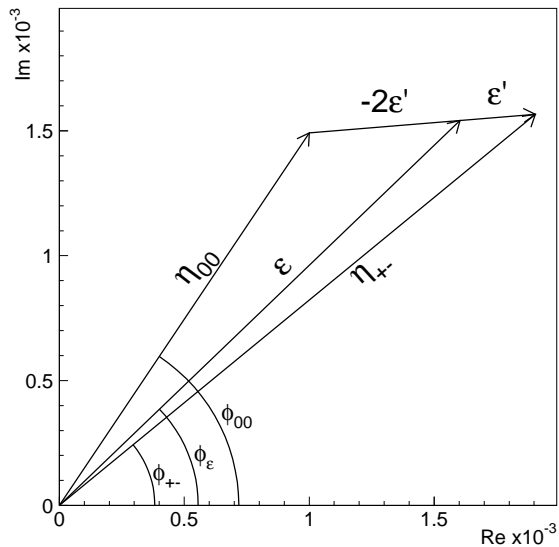,width=0.9\linewidth}
\end{center}
\caption{\label{fig:kaonparams}
Diagram of CP violating kaon parameters.  For this illustration, the
$\epsilon$ parameter has the central value measured by KTeV and
the value of $\epsilon'$ is scaled by a factor of 50. Although they
appear distinct in this diagram,
note that $\phipm$ and 
$\phizz$
are consistent with each other within experimental errors.}
\end{figure}

Since the 1964 discovery of CP violation in $\KLpm$ decay\cite{ccft64},
significant experimental effort has been devoted to understanding the mechanism
of CP violation. Early experiments showed that the observed effect was due
mostly to a small asymmetry between
the $\Kz \to \Kzbar$ and $\Kzbar \longrightarrow \Kz$
transition rates,
which is referred to as indirect CP violation. Decades of additional effort
were required to demonstrate the existence of direct CP violation in a decay
amplitude.
This paper reports the final measurement
of direct CP violation by the KTeV Experiment (E832) at Fermilab.

Direct CP violation can be detected by comparing the level of CP
violation for different decay modes.  The parameters $\epsilon$ and
$\epsilon'$ are related to the ratio of CP violating to CP conserving
decay amplitudes for $\Kpm$ and $\Kzz$:
\begin{equation}
 \begin{array}{lcccc}
  \etapm & \equiv & \frac{\textstyle A \left( \KLpm\right)}
                    {\textstyle A \!\left(\KSpm\right)} &
  \approx & \epsilon + \epspr, \\
  \etazz & \equiv &\frac{\textstyle A \left( \KLzz\right)}
                   {\textstyle A \!\left(\KSzz\right)} & \approx
  & \epsilon -  2\epspr,
 \end{array}
\end{equation}
where $\epsilon$ is a measure of indirect CP violation, which is common
to all decay modes. The relation among the complex parameters $\etapm$, $\etazz$,
$\epsilon$, and $\epsilon'$ is illustrated in Fig.~\ref{fig:kaonparams}.

If CPT symmetry holds, the phase of $\epsilon$
is equal to the ``superweak'' phase:
\begin{equation}
  \phisw \equiv \tan^{-1} \left( 2 \delm / \Delta \Gamma \right),
\end{equation}
where $\delm\equiv m_L - m_S$ is the $K_L$-$K_S$ mass difference and
$\Delta\Gamma =\Gamma_S - \Gamma_L$ is the difference in the
decay widths.

The quantity $\epsilon^{\prime}$ is a measure of direct CP violation,
which contributes differently to the $\pi^+\pi^-$ and $\pi^0\pi^0$ decay
modes,
and is proportional to the difference between the decay amplitudes
for $\Kz\to\pi^+\pi^-(\pi^0\pi^0)$ and
$\Kzbar\to\pi^+\pi^-(\pi^0\pi^0)$.
Measurements of $\pi\pi$ phase shifts~\cite{ochs} show that,
in the absence of CPT violation, the phase of $\epsilon'$ is
approximately equal to that of $\epsilon$. Therefore, $\reepoe $ is a
measure of direct CP violation and $\imepoe $ is a measure of CPT
violation.

Experimentally,  $\reepoe$ is determined from the double ratio of the
two pion decay rates of $K_L$ and $K_S$:
\begin{eqnarray}
  \frac{\Gamma\!\left(\KLpm\right)/\,\Gamma\!\left(\KSpm\right)}{
        \Gamma\!\left(\KLzz\right)/\,\Gamma\!\left(\KSzz\right)}
                  \nonumber \\
        = \left| \frac{\etapm}{\etazz} \right|^2
        \approx  1 + 6 \reepoe.
              &  &
  \label{eq:reepoe}
\end{eqnarray}
For small $|\epsilon'/\epsilon|$, $\imepoe$ is related
to the phases of $\etapm$ and $\etazz$ by
\begin{equation}
\begin{array}{lcl}
  \phipm &\approx &\phiep  + \imepoe,  \\
  \phizz &\approx &\phiep  - 2 \imepoe,  \\
  \delphi &\equiv &\phizz  - \phipm   \approx -3  \imepoe~.
\end{array}
  \label{eq:delphimpe}
\end{equation}

The Standard Model accommodates both direct and indirect
CP violation \cite{ckm,ellis,gilman_wise}.
Most recent Standard Model predictions
for $\reepoe$ are less than $30 \times 10^{-4}$
\cite{kaon99:ciuchini,hepph:belkov,jhep:bijnens,nbp:hambye,Pallante:2001he,
prd:wu,npb:narison,thesis:sanchez,Buras:2003zz,pich04};
however, there are large hadronic uncertainties in these calculations.
Experimental results have established that $\reepoe$ is non-zero
\cite{prl:731,pl:na31,prl:pss,na48:reepoe,prd03}.
The previous result from KTeV, which was based on about half of the KTeV
dataset, is $\reepoe = (20.7 \pm 2.8)\eu$\cite{prd03}.  This result was published
in 2003 and will be referred to in this text as ``KTeV03.''
The result based on all data from NA48 at CERN is 
$\reepoe = (14.7 \pm 2.2)\eu$\cite{na48:reepoe}.

This paper reports the final measurement of $\reepoe$ by KTeV.
The measurement is based on 85 million reconstructed \Kpp\ decays
collected in 1996, 1997, and 1999.
This full sample is two 
times larger than, and contains, the sample on which the KTeV03
results are based.
We also present measurements of
the kaon parameters $\delm$ and $\tauS$,
and tests of CPT symmetry based on measurements of
$\delphi$ and $\phiep -\phisw$.  Using our phase measurements,
we place a limit on the mass difference between
$\Kz$ and $\Kzbar$.

For this analysis we have made significant improvements to the data 
analysis and the Monte Carlo simulation.
The full dataset, including the data used in KTeV03,
has been reanalyzed using the improved reconstruction and simulation.
These results supersede the previously published KTeV03 results\cite{prd03}, 
which were based on data from 1996 and 1997.  

This paper describes the KTeV experiment in Sec. \ref{sect:expt},
the analysis technique in Sec. \ref{sect:ana}, and the extraction of 
physics results in Sec. \ref{sect:accandfit}.  
We emphasize changes and improvements since the KTeV03 publication.  
We will refer to \cite{prd03} for some details that have not changed
since KTeV03.
Section \ref{sect:results} presents the final KTeV results, including
correlations between the parameters and crosschecks of the results.
Section \ref{sect:conclude} is a summary and discussion of the results.
Appendix \ref{sec:regeneration}
contains a discussion of the dependence of our measurements on details
of kaon regeneration.

\section{\label{sect:expt}Measurement Technique and Apparatus}

\subsection{Overview}
The measurement of \reepoe\ requires a source of $K_L$ and $K_S$
decays, and a detector to reconstruct the charged ($\ppc$)
and neutral ($\ppn$) final states.
The strategy of the \ktev\ experiment is to produce two identical
$K_L$ beams, and then to pass one of the beams through a
``regenerator'' that is about two hadronic interaction lengths long.
The beam that passes through the regenerator is called the
regenerator beam,
and the other beam is called the vacuum beam.
The regenerator creates a coherent
$\ket{K_L}+\rho\ket{K_S}$ state,
where $\rho$, the regeneration amplitude, is a physical property
of the regenerator.  The regenerator is designed
such that most of the \Kpp\ decays
downstream of the regenerator are from the $K_S$ component.
The charged spectrometer is the 
primary detector for reconstructing $\Kpm$ decays and the 
pure Cesium Iodide (CsI) calorimeter 
is used to reconstruct the four photons from $\Kzz$ decays.  
A Monte Carlo simulation is used to correct for the average acceptance
difference between \Kpp\ decays in the two beams,
which results from the very different $K_L$ and $K_S$ lifetimes.
The decay-vertex distributions provide a
critical check of the simulation.
The measured quantities are the vacuum-to-regenerator
``single ratios''  for \Kpm\ and \Kzz\ decay rates.
These single ratios are proportional to
$|\etapm/\rho|^2$ and $|\etazz/\rho|^2$, respectively, 
and the ratio of these two quantities gives $\reepoe$
via Eq.~\ref{eq:reepoe}.

\subsection{KTeV Experiment}
The KTeV kaon beams are produced by a beamline of magnets, absorbers, 
and collimators that act on the products of a proton beam 
incident on a fixed target.  The 800 GeV/$c$ proton beam, provided 
by the Fermilab 
Tevatron, has a 53 MHz RF structure so that the protons arrive in $\sim$1 ns
wide ``buckets'' at 19 ns intervals.  This beam is incident on a beryllium oxide 
(BeO) target that is about 
one proton interaction length long.
Immediately downstream of the target, the beam consists of protons, muons, 
and other charged particles, neutral kaons, neutrons, photons, and hyperons.  
This beam is collimated into two beams and the non-kaon component is reduced
by magnets and absorbers in a 100 meter long beamline.
At the start of the fiducial decay region, 120 m downstream of the target,
the average kaon momentum is about $70$ GeV/$c$.  
The neutron-to-kaon
ratio is $1.3$ in the vacuum beam and $0.8$ in the regenerator beam. The KTeV
beams and the beamline elements that produce them are described in detail 
in \cite{prd03}.

\begin{figure}
\begin{center}
\epsfig{file=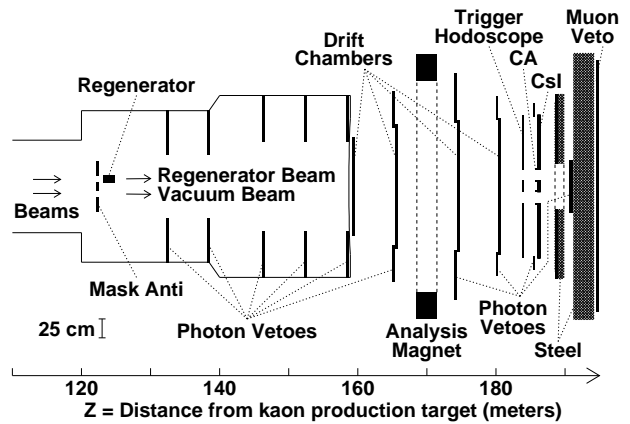,width=\linewidth}
\caption{Schematic of the KTeV detector. Note that the vertical and horizontal
scales are different.}
\label{fig:detector}
\end{center}
\end{figure}

KTeV reconstructs kaon decays that occur in an evacuated decay region 90-160 m
downstream of the target.  Figure \ref{fig:detector} is a 
schematic of the detector.  In the KTeV coordinate system, the
positive $x$-axis points to the left if the observer is facing downstream,
the positive $y$-axis points up, and the positive $z$-axis points downstream
from the target.
At the upstream end of the decay region,  
the regenerator alternates between the 
two beams 
to minimize acceptance differences between decays in the vacuum and regenerator beams.  
The
charged spectrometer and CsI calorimeter are located downstream of the
vacuum window at the end of the decay region. The decay region and
primary detectors are surrounded by a system of photon veto detectors to detect
particles with trajectories that miss the CsI calorimeter.  The major detector
elements are described in more detail in the following paragraphs. 

\begin{figure}
  \epsfig{file=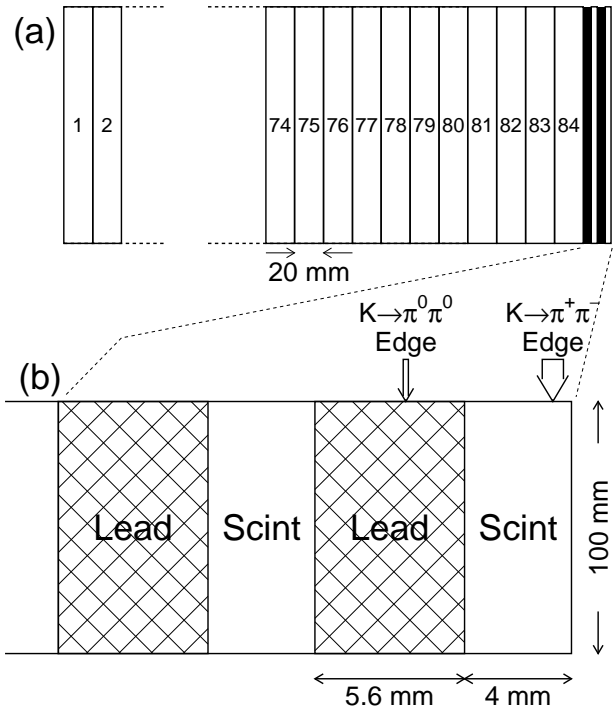,width=\linewidth}
  \caption{
      Diagram of the regenerator.  (a) Layout of the 85
      regenerator modules, including the lead-scintillator
      module.  (b) Zoomed diagram of the lead-scintillator
      regenerator module.
      The PMTs above and below are not shown.
      The thickness of each lead (scintillator) piece is
      5.6 (4.0) mm.  The transverse dimension is 100~mm,
      and is not drawn with the same scale as the $z$-axis.
      The kaon beam enters from the left.
      The arrows indicate the location and $\pm 1\sigma$
      uncertainty of the effective upstream edges for 
      reconstructed \Kzz\ and \Kpm\ decays
      for 1999 data.
           }
  \label{fig:regedge}
\end{figure}

%The regenerator is active and is used as part of the trigger and veto systems 
%as well as to provide $\KS$ regeneration.  
%It 
The regenerator
consists mainly of 84 ~$10\times10\times 2~{\rm cm}^3$ scintillator
modules as seen in Fig. \ref{fig:regedge}a.
Its primary purpose is to provide $\KS$ regeneration,
but it is also used as part of the trigger and veto systems.
Each module is viewed by two photomultiplier tubes (PMTs),
one from above and one from below.
The downstream end of the regenerator has a lead-scintillator sandwich
called the ``regenerator Pb module'' (Fig.~\ref{fig:regedge}b),
which is also viewed by two PMTs.
This last module of the regenerator is used to define a sharp
upstream edge for the kaon decay region in the regenerator beam.

\begin{figure}
  \centering
  \epsfig{file=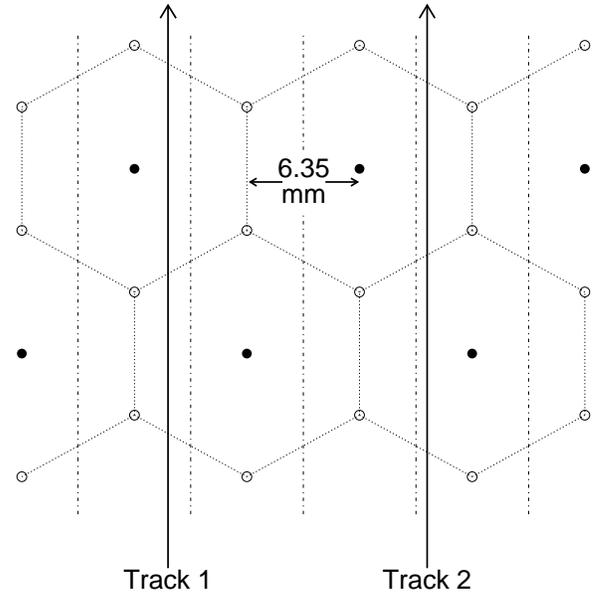,width=\linewidth}
  \caption{
      Diagram of drift chamber geometry showing six field wires
      (open circles) around each sense wire (solid dots). The solid
      lines illustrate the hexagonal cell geometry; they do not represent
      any physical detector element.  The vertical dashed lines are separated
      by 6.35 mm and are used
      to define the track separation cut described in Sec. \ref{sect:chsel}.
        }
  \label{fig:dccell}
\end{figure}

The charged spectrometer 
%is the 
%primary detector for reconstructing $\Kpm$.  It
consists of four drift chambers (DCs) and a dipole magnet.
Each drift chamber measures charged-particle
positions in both the $x$ and $y$ views.
A chamber consists of two planes of horizontal wires
to measure $y$ hit coordinates,
and two planes of vertical wires to measure $x$ hit coordinates;
the two $x$-planes and the two $y$-planes are offset to resolve
position ambiguities.
The DC planes have a hexagonal cell geometry formed by six
field-shaping wires surrounding one sense wire
(Fig.~\ref{fig:dccell}).  There are a total of 1972 sense wires
in the four drift chambers.
The cells are $6.35$~mm wide,
and the drift velocity is about $50 \mu$m/ns.
The analyzing magnet imparts a kick of 412 MeV/$c$ in
the horizonal plane.
The well-known kaon mass is used to set the momentum scale
with $10^{-4}$ precision.

\begin{figure}
    \centering
    \epsfig{file=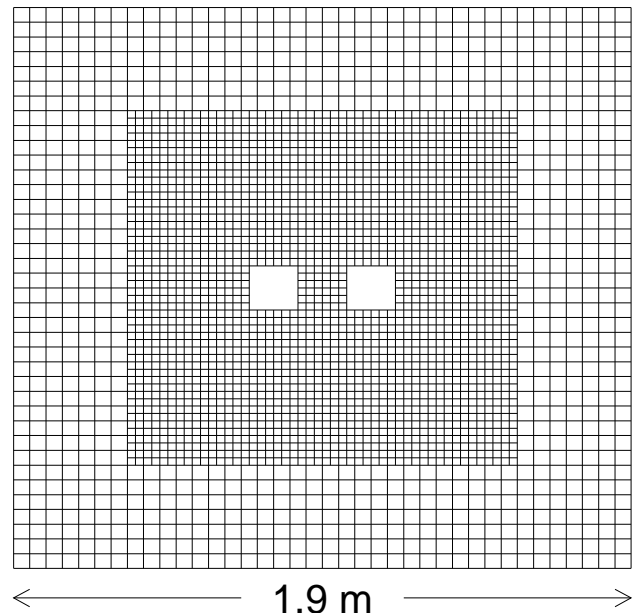,width=\linewidth}
    \caption{
        Beamline view of the KTeV CsI calorimeter,
        showing the 868 larger outer
        crystals and the 2232 smaller inner crystals.
        Each beam hole size is  $15\times 15$~cm$^2$ and
        the two beam hole centers are separated by 0.3~m.
        The positive $z$ direction is into the page.
           }
    \label{fig:csilayout}
  \end{figure}

The CsI calorimeter 
%is used to reconstruct the four photons from $\Kzz$ decays.  It 
consists of 3100 pure CsI
crystals viewed by photomultiplier tubes.
The layout of the $1.9 \times 1.9$~m$^2$ calorimeter is shown in
Fig.~\ref{fig:csilayout}.
There are 2232 $2.5 \times 2.5$ cm$^2$ crystals in the central region,
and 868 $5 \times 5$ cm$^2$ crystals surrounding the smaller crystals.
The crystals are all 50~cm (27 radiation lengths) long.  Each crystal
is wrapped in 12 $\mu$m, partially-blackened, aluminized mylar in a manner 
designed to make the longitudinal
response of each crystal as uniform as possible. The calorimeter is
read out by custom digitizing electronics (DPMTs) placed directly behind the
PMTs\cite{dpmt}.  Momentum-analyzed electrons and positrons from 
$\ke3$ ($\Kethree$) decays are used to calibrate the CsI energy scale 
to 0.02\%.

An extensive veto system is used to reject events coming from
interactions in the regenerator,
and to reduce background from kaon decays into
non-$\pi\pi$ final states
such as $K_L\to\kmth$ and $\Kzzz$.
The veto system consists of a number 
of lead-scintillator detectors in and around the primary detectors.  

KTeV uses a three-level trigger to select events.
Level~1 uses fast signals 
from the detector and introduces no deadtime.  Level 2 is based on
more sophisticated processing from custom electronics and introduces
a deadtime of 2-3 $\mu$s; when an event passes Level 2 the entire detector
is read out with an average deadtime of 15 $\mu$s.  Level 3 
is a software filter; the processors have enough memory that no further
deadtime is introduced.   

Individual triggers are defined to select $\kchrg$ and $\kneut$ decays;
the trigger efficiencies are studied using decays collected in separate
minimum-bias triggers.  Additional triggers select decays such as $\ke3$
and $\Kzzz$ which are used for calibration and acceptance studies.  The
``accidental'' trigger uses a set of counters near the target to collect
events based on primary beam activity; these events are uncorrelated with 
detector signals that come from the beam particles and are used to model the 
effects of intensity-dependent accidental activity.

Several changes were made to the KTeV experiment to improve data collection
efficiency for the 1999 run.

\begin{enumerate}

\item \emph{Neutral Beams.} The proton extraction cycle of the Tevatron 
was improved from 20 second 
extractions (or ``spills'')
every 60 seconds in 1996 and 1997 to 40 second extractions every 80 seconds 
in 1999.  The maximum available intensity was $\sim$2 $\times 10^{11}$ 
protons per second.  In 1999, KTeV chose to take 
about half of the data at 
an average intensity of $\sim 1.6 \times 10^{11}$ protons/s and half
at a lower average intensity of $\sim 1 \times 10^{11}$ protons/s 
%high intensity and half at about twice 
%lower intensity 
as a systematic cross-check.  

\item \emph{CsI Calorimeter Electronics.} During 1996 and 1997 data taking, 
individual channels of the custom 
readout electronics for the CsI calorimeter failed occasionally.  These failures 
account for half 
of the 20\% data-taking inefficiency during 1996 and 1997.  They also affect 
the data quality and complicate the calibration of the calorimeter.  
All of the custom electronics were 
re-fabricated and installed in the CsI calorimeter in preparation for the 1999 run.  
The re-fabrication of the chips was successful; no CsI calorimeter electronics had to 
be replaced during the 1999 run.

\item \emph{Drift Chambers.} The drift chambers required some repair 
due to radiation damage sustained during data taking in 1996 and 1997.
About half of one drift chamber was restrung and a second chamber 
was cleaned.  
The drift chamber readout electronics were modified to allow the system
to run at higher gain without causing the system to oscillate or trigger
on noise.  

\item \emph{Helium Bags.} Helium bags are placed between the drift 
chambers to minimize the matter 
seen by the neutral beams after leaving the 
vacuum decay region and to reduce multiple scattering of charged
particles.  In 1996 and 1997, one of the small helium bags was leaky and 
contained mostly air by the end of the 
1997 run, so it was necessary in the analysis to correct for the
increased multiple scattering resulting from this increased material in
the detector.  The bags were replaced for 1999.  This
reduction in material traversed by the beams was offset by 
a change in the buffer gas used in the drift chambers; 
the total ionization loss upstream of the
CsI calorimeter was less than 5 MeV in each year.
This energy loss occurs mostly in the scintillator hodoscope 
just upstream of the CsI calorimeter.

\item \emph{Trigger.} In 1999, the trigger was adjusted to select more 
$\KLpmz$ decays for the measurement of kaon flux attenuation in the
regenerator beam, called ``regenerator transmission.'' 
The improvement of this measurement reduces
several systematic uncertainties associated with the fitting procedure
as described in Sec. \ref{sect:fit}.

\end{enumerate}

\subsection{\label{sect:mcsim}Monte Carlo Simulation}
KTeV uses a Monte Carlo (MC) simulation to calculate the detector
acceptance and to model background to the signal modes.  
The different $\KL$ and $\KS$ lifetimes lead to different 
$z$-vertex distributions in the 
vacuum and regenerator beams.  We determine the 
detector acceptance as a function of kaon decay $z$-vertex and energy, including 
the effects of geometry, detector response, and resolution.  
To help verify the accuracy of the MC simulation, we collect 
and study decay modes with approximately ten times higher statistics than 
the $\Kpp$ signal samples: $\ke3$, $\Kzzz$, and $\KLpmz$.

The Monte 
Carlo simulates $\Kz$/$\Kzbar$ generation at the BeO target following the
parameterization in \cite{malensek}, propagates the coherent $\Kz$/$\Kzbar$
state through the absorbers and collimators along the beamline to the decay point,
simulates the decay including decays inside the regenerator, 
traces the decay products through the detector,
and simulates the detector response including the digitization of the 
detector signals and the trigger selection.  The parameters of the detector geometry are 
based both
on data and survey measurements.
Many aspects of the 
tracing and detector response are based on samples of detector responses, 
called ``libraries,'' that are generated with 
GEANT\cite{GEANT} simulations; the use of libraries keeps the MC
relatively fast.
The effects of accidental activity are included in the simulation by 
overlaying data events from the accidental trigger onto the simulated events.  
The Monte Carlo event format is identical to data and the events are 
reconstructed and analyzed in the same manner as data.
% with very few exceptions.  
More details of the simulation are available in \cite{prd03}.

Many improvements have been made to the MC simulation since 
KTeV03\cite{prd03}.  We have improved
the simulation to include finer details of electromagnetic showering
in the CsI calorimeter and charged particle propagation through the detector.  
These changes are described in detail below.

\begin{enumerate}

\item \emph{Shower library.} For this analysis, the GEANT-based 
library used to simulate 
photon and electron showers in the CsI calorimeter has been improved to 
simulate 
the effects of incident particle angle.  The library used for KTeV03
 was binned in energy and incident position.  There were 325 position
bins and six logarithmic energy bins (2 GeV, 4 GeV, 8 GeV, 16 GeV, 32 GeV,
and 64 GeV).  The effect of angles 
was approximated by shifting the incident position based on the angle of 
incidence.  The shower library has now been expanded to include nine angles in $x$ and $y$
(-35 mrad to 35 mrad) for photons and 15 angles in $x$ and $y$ (-85 mrad to 85 mrad) for 
electrons.  Electron angles may be larger than photon angles because of
the momentum kick imparted by the analyzing magnet.  The position and energy binning
is unchanged from KTeV03.  Differences between 
the library angle and the desired angle are 
accounted for by shifting the incident position.
The particle energy cutoff applied 
in the GEANT shower library generation
has been lowered from 600 keV to 50 keV for electrons; the photon
cutoff of 50 keV is unchanged.
Sixteen showers per bin have been generated.  
Energy deposits are corrected for energy lost in the 12 $\mu$m mylar
wrapping around the CsI crystals.

The current Monte Carlo produces a significantly better 
simulation of electromagnetic showers in the CsI calorimeter.  
Figure \ref{fig:ke3shwr2}
shows the data-MC comparison of the fraction of energy 
in each of the 49 CsI crystals in electron
showers relative to the total reconstructed shower energies for 
electrons from $\ke3$ decays.  The majority of the energy is deposited in the
central crystal since the Moliere radius of CsI is 3.8 cm.  
These plots are made for 16-32 GeV electrons with incident angles 
of 20-30 mrad; the quality of 
agreement is similar for other energies and angles.  The data-MC agreement 
improves dramatically as seen in Fig. \ref{fig:ke3shwr2}.  This improvement in the
modeling of electromagnetic shower shapes leads to important reductions 
in the systematic
uncertainties associcated with the reconstruction of photon showers from
$\kneut$ decays (see Sec. \ref{sect:ana_neut}).

\begin{figure}
\begin{center}
\epsfig{file=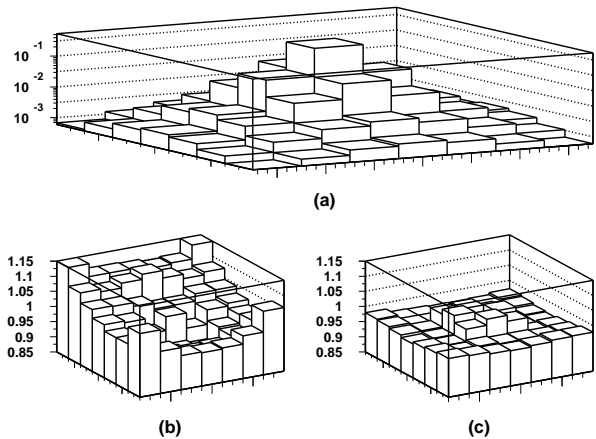,width=\linewidth}
\caption{Data-MC comparison of fraction of energy in each of the 49 CsI
crystals in an electron shower.  
(a) The fraction of energy in each of the 49 CsI crystals in electron
showers for data.  (b) KTeV03 data/MC ratio. (c) Current data/MC ratio.}
\label{fig:ke3shwr2}
\end{center}
\end{figure}

\item \emph{Ionization Energy Loss.} In KTeV03,
we did not include the effect of 
ionization energy losses for charged particles in the simulation.  
In the current simulation, we include the ionization loss for each volume 
of material in the detector.  The total loss up to the surface of the CsI is 
less than 5~MeV.  This is 
a very small effect for $\kchrg$ decays but it is important for low-energy 
electrons used in the calibration of the CsI calorimeter and affects
converted photons from $\kneut$ decays.

\item \emph{Bremsstrahlung.} In KTeV03, the MC included 
electron Bremsstrahlung 
in materials upstream of the analyzing magnet only.  In the current analysis, 
the Bremsstrahlung rate and photon angle in each volume of material in the detector are
included in the simulation. 

\item \emph{Delta Rays.} In the KTeV03 simulation, delta rays produced 
in a drift chamber cell deposited all of their energy in that cell.  The MC 
now has a more complete treatment 
in which delta rays may scatter into adjacent cells of the drift chamber.
High momentum delta rays are traced through the detector like any 
other particle and low momentum delta rays are simulated using a library
created with GEANT4\cite{GEANT4}.  This treatment of delta rays improves 
our simulation of the distribution of extra in-time hits in the drift 
chambers.

\item \emph{Pion Interactions.} The probability for pions to interact 
hadronically with material in the spectrometer
is $0.7\%$; hadronic interactions in the spectrometer were not 
simulated in KTeV03.
These events
are now simulated using a GEANT-based 
library which contains a list of secondary particles produced by each hadronic 
interaction.  An average of nine secondary particles are produced per 
interaction; these secondary particles are read in from the shower
library and traced through the rest of the detector like any other particle.  
Events with pion interactions typically trigger the photon veto system 
and so do not pass selection criteria in 
the analysis.

\item \emph{Fringe Fields.} The simulation of fringe fields from the 
analysis magnet has been refined.
Fringe fields from the analysis magnet inside the vacuum tank and between
all four drift chambers have been measured and are now simulated.  The 
maximum strength of the fringe field in these regions is comparable to the earth's magnetic field.  
The fringe field simulation improves the MC description of the 
azimuthal dependance of the $\pi^+\pi^-$ invariant mass 
for $\kchrg$ data.

\item \emph{Position Resolution.} The position resolution of the drift 
chambers is dependent upon position 
within the cell as shown in Fig. \ref{fig:chrgres}.  
In KTeV03, the resolution was treated as 
flat across the cell; the position dependence of the resolution is now 
simulated.  The position dependence of the resolution is also considered
in the analysis of $\kchrg$ data; this is described in Sec. \ref{sect:ana_chrg}.

\begin{figure}
\begin{center}
\epsfig{file=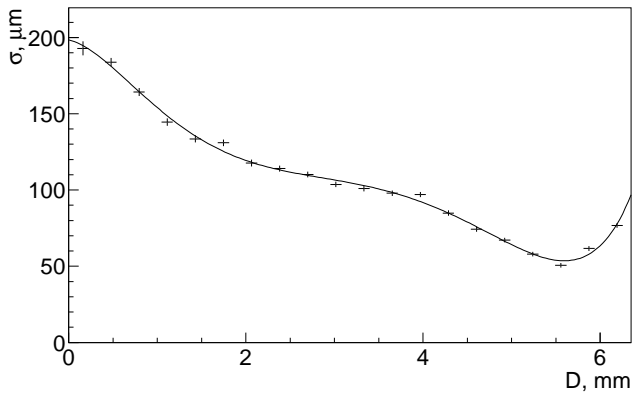,width=\linewidth}
\caption{Dependence of drift chamber position resolution on position
within the cell.  D is the distance from the sense wire.
Crosses represent the measured central values and uncertainties
of the resolution in bins of $D$ and the line represents a polynomial
fit to the data.  This fit is used in the simulation to parameterize the 
position dependence of the position resolution. }
\label{fig:chrgres}
\end{center}
\end{figure}

The position resolution of CsI calorimeter clusters in the MC is slightly worse than 
in data.  To better
match the resolutions, we artificially improve the resolution of the MC 
by 9\%.
This is done by moving the reconstructed position toward the generated 
particle position. 

%\item \emph{Input Parameters.} We now use the following
%kaon parameters as inputs to the simulation:
%\begin{itemize}
%\item \tauL = 5.097 $\times 10^{-8}$ s
%\item \tauS = 8.9619 $\times 10^{-11}$ s
%\item $\vert\epsilon\vert$ = 0.002224
%\item $m_{K}$ = 497.626 $\umass$.
%\end{itemize}

\end{enumerate}

\section{\label{sect:ana}Data Analysis}
The analysis is designed to identify \Kpp\ decays while removing poorly 
reconstructed events that are difficult to simulate, and to reject background.
For each decay mode, the same requirements are applied to
decays in the vacuum and regenerator beams,
so that most systematic uncertainties cancel in
the single ratios used to measure
$|\etapm/\rho|^2$ and $|\etazz/\rho|^2$.
The following sections describe the analysis and the
associated systematic uncertainties in $\reepoe$.

\subsection{Common Features}
Although many details of the charged and neutral decay mode
analyses are different, several features are common
to reduce systematic uncertainties.
We select an identical $40$-$160~\upk$ kaon momentum range for
both the charged and neutral decay modes.
We also use the same $z$-vertex range of 110-158~m from the target
for each decay mode.
To simplify the treatment of background from kaons that
scatter in the regenerator, the veto requirements for the
charged and neutral mode analyses
are made as similar as possible.  

When discussing systematic uncertainties, we typically estimate a
potential shift $s \pm \sigma_s$, where $s$ is the shift in $\reepoe$
and $\sigma_s$ is the statistical uncertainty on $s$.  We assign a
symmetric systematic error, $\Delta_s$, such that the range 
$[-\Delta_s,+\Delta_s]$ includes 68.3\% of the area of a Gaussian with 
mean $s$ and width $\sigma_s$.

\subsection{\label{sect:ana_chrg}Charged Reconstruction and Systematics}

The $\kchrg$ analysis consists primarily of the reconstruction of tracks 
in the spectrometer; the vertices and momenta of the tracks are used to 
calculate kinematic quantities describing the decay.  
The CsI calorimeter is used for particle identification by comparing
the reconstructed cluster energy to the measured track momentum.
The analysis requirements provide clean identification of well-reconstructed 
$\kchrg$ events
with little background contamination. The cuts are sufficiently loose
to reduce systematics from modeling of resolution tails.
The $\kchrg$ reconstruction and event selection are described in the
following
sections; more details of the analysis are found
in \cite{prd03}.

\subsubsection{$\kchrg$ Reconstruction}

The spectrometer reconstruction begins by finding
tracks separately in the $x$- and $y$-views.
Track segments are found in the two drift chambers
upstream of the magnet and the two drift chambers
downstream of the magnet; these segments are then
extrapolated to the center of the magnet. 
We require the extrapolated segments to match within 6~mm at the
magnet mid-plane to form a combined track; they typically match to within 0.5~mm.
Each particle momentum is determined from the track bend-angle
in the magnet and a map of the magnetic field.  

The process of finding track segments depends on the alignment and
calibration of the drift chambers.
For the current analysis, we made new measurements of the 
drift chamber sizes and rotations.
The survey of the wire positions used a large coordinate measurement 
machine with a camera and magnifying lens mounted on the end of a movable arm.
The measured drift chamber size is about 0.02\% larger than the nominal 
value found by scaling the 6.35 mm ``cell'' size.  
The relative non-orthogonality
between DC1 and DC2
is limited to $\pm$30 $\mu$rad.  The uncertainty in $\reepoe$ associated
with the drift chamber alignment and calibration is 0.20$\eu$.  The momentum
measurement uses the known kaon mass  as a constraint; the 0.022 MeV/$c^2$ 
uncertainty in the kaon mass corresponds to an uncertainty in $\reepoe$ of
0.10$\eu$.

If two $x$-tracks and two $y$-tracks are found,
the reconstruction continues by extrapolating both sets
of tracks upstream to define
vertices projected on the $x$-$z$ and $y$-$z$ planes.
The difference between these two projections, $\dzvtx$,
is used to define a vertex-$\chi^2$,
\begin{equation}
   \chisqvtx \equiv \left( {\dzvtx}/\sigdz \right)^2 ~,
   \label{eq:chisqvtx}
\end{equation}
where $\sigdz$ is the resolution of $\dzvtx$.
This resolution depends on momentum and opening angle, 
and accounts for multiple scattering effects.
The two $x$-tracks and two $y$-tracks are assumed to originate
from a common vertex if $\chisqvtx < 100$.

To determine the full particle trajectory,
the $x$ and $y$ tracks are matched to each other
based on their projections to the CsI;
the projected track positions must match CsI
cluster positions to within 7~cm.

An event is assigned to the regenerator beam if the
regenerator $x$-position has the same sign as
the $x$-coordinate of the kaon trajectory 
at the downstream face of the regenerator;
otherwise, the event is assigned to the vacuum beam.

In KTeV03~\cite{prd03}, the track segments
were reconstructed assuming that the position resolution of the drift
chambers does not depend on the hit position 
within a chamber cell. To check this assumption, a special data 
sample was collected
with the  magnetic field turned off. 
Three chambers are used to reconstruct straight tracks and
these tracks are
compared to the hits reconstructed in the fourth chamber.
The resulting position resolution (see Fig.~\ref{fig:chrgres})
shows a significant dependence on the distance between the track 
and the sense wire.
Tracks passing close to a sense wire have worse resolution because
of the time separation of drift electrons reaching the sense
wire.
The variation of the resolution for 
tracks close to the boundary of a drift cell can be partly explained
by the electric field configuration. 
The drift-time dependent resolution
is included in the MC and is used for the track segment reconstruction 
in the current analysis. The new resolution measurement improves 
reconstruction of $z$-vertex and track momentum.  For example,
the width  of  the $\pi^+\pi^-$ invariant mass is reduced by $\sim 14\%$;
this improvement is more significant for higher momentum kaons where
it reaches $\sim 25\%$.
 
The kaon decay vertex position and the momenta of the two tracks 
forming the vertex are used to calculate the $\pppm$ invariant mass,
their energy, and $\ptsq$, the sum of their momenta transverse to the
beam direction.

\subsubsection{\label{sect:chsel}$\kchrg$ Selection}
The $\kchrg$ event selection begins with the three-level trigger during data
taking.  Level 1 uses hits in the trigger hodoscopes and the drift
chambers to select events consistent with two charged particles coming
from the decay of a kaon
that did not undergo large angle scattering 
in the defining collimator or regenerator
prior to the decay.  
Level 2 uses custom hit counting
electronics and a track finding system to select events with two tracks
from a common vertex.  The vertex requirement at trigger level is loose
compared to the selection criteria in the offline analysis.
 The inefficiencies of the Level 1 and Level 2 
triggers are studied 
using $\ke3$ decays from minimum-bias triggers.  The uncertainty in $\reepoe$
associated with Level 1 and Level 2 event selection is 0.2$\eu$.  
The Level 3 software filter reconstructs two charged tracks and makes
loose cuts on reconstructed mass and particle indentification.  
To measure the Level 3 inefficiency of the $\kchrg$ trigger, we 
perform the full offline analysis on ``random
accepts,'' a prescaled subset of 
the $\kchrg$ trigger that has no Level 3
requirement.   
We find that the bias in $\reepoe$ from the Level 3 trigger
inefficiency is (0.30 $\pm$ 0.12)$\eu$.  We correct $\reepoe$ for this
bias and assign an uncertainty in $\reepoe$ of 0.12$\eu$.

The offline selection criteria for $\kchrg$ decays are tighter than those
imposed by the trigger. The $\kchrg$ analysis requirements and any
associated systematic uncertainties in $\reepoe$ are described in 
the following paragraphs.

We make a number of cuts on energy deposits in the veto detectors.  
The most important veto requirement are the muon veto cuts, which suppress 
background from $\KLmu$ decays, and the regenerator cuts, which reduce 
background from scattered kaons.  Additional veto cuts are made for 
consistency with the $\kneut$ analysis.  

We also use the spectrometer and the calorimeter as ``veto detectors.''
We reject events with any tracks other than those from the vertex.
We require the ratio of reconstructed cluster energy to track momentum,
$E/p$, to be less than 0.85 to 
identify the tracks as pions. We require that
the track momentum be greater than 8~GeV/$c$ to ensure $100\%$ efficiency 
for the muon veto detectors.
These cuts suppress background from
$\Kethree$ and $\Kmuthree$ decay modes.  

We remove events with $1.112$~GeV/$c^2$ $<$ $m_{p\pi}$ $<$ $1.119$~GeV/$c^2$,
where $m_{p\pi}$ is the reconstructed invariant mass assuming the higher 
momentum particle is a proton.  This removes
background from $\Lppi$ and $\bar{\Lambda}\to\bar{p}\pi^+$ decays where the proton is mis-identified as a pion.  

\begin{figure}
\begin{center}
\epsfig{file=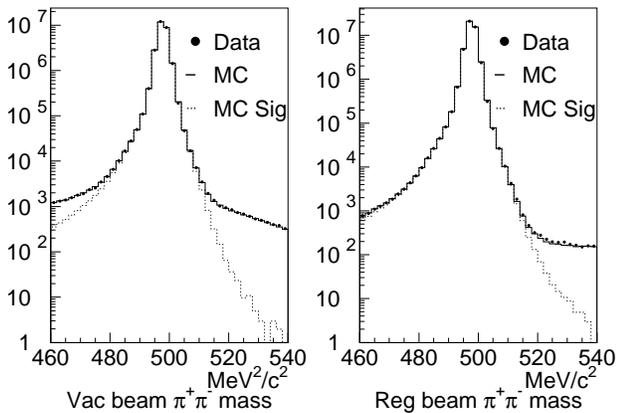,width=\linewidth}
\end{center}
\caption{\label{fig:massch}
$\pi^+\pi^-$ invariant mass distribution for $\kchrg$ candidate
events in the vacuum (left) and regenerator (right) beams. 
The data distribution is shown as dots, the $\kchrg(\gamma)$ 
signal MC (MC Sig) is shown as dotted histogram and the sum
of signal and background MC is shown as a solid histogram.}
\end{figure}

\begin{figure}
\begin{center}
\epsfig{file=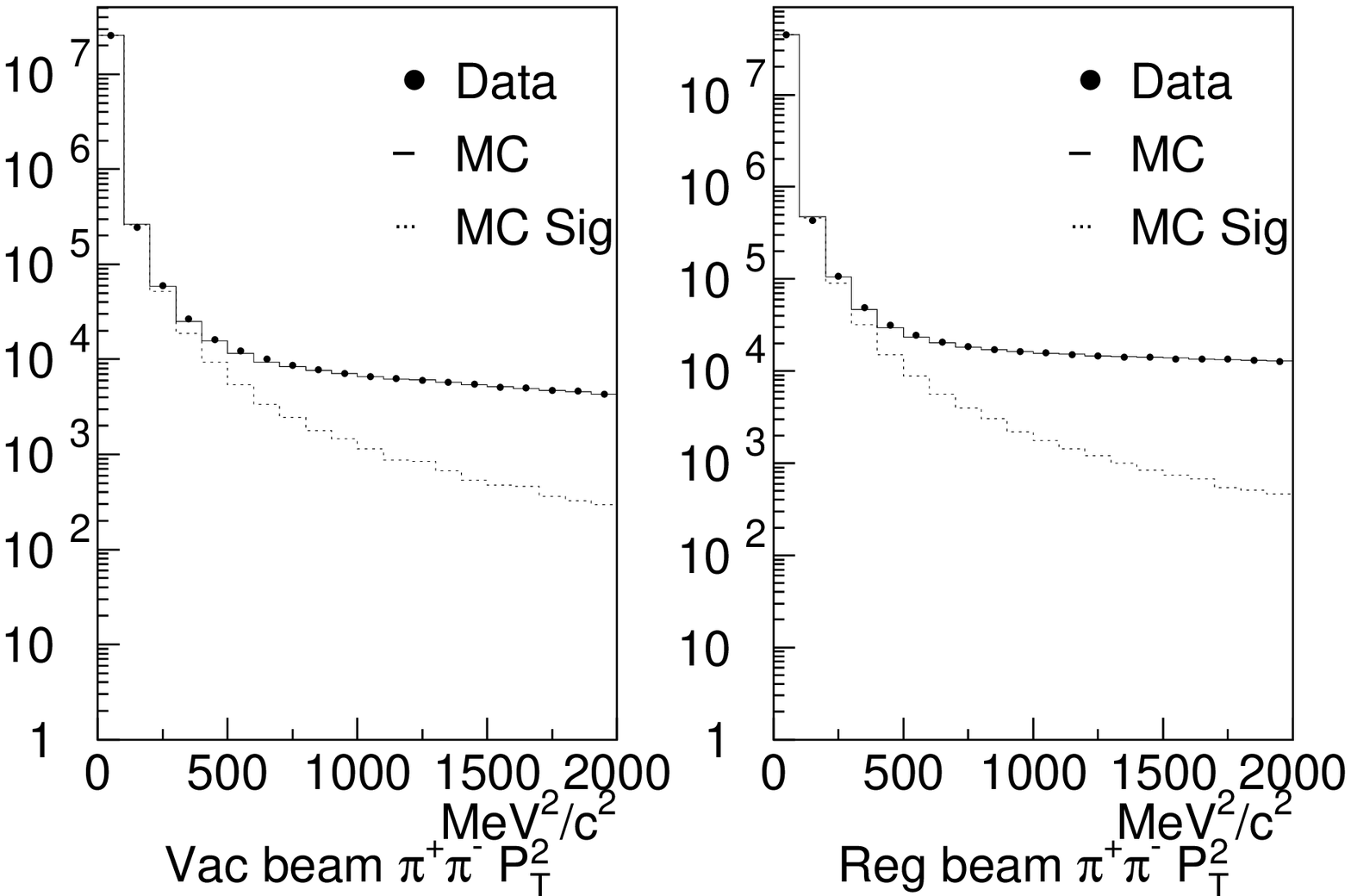,width=\linewidth}
\end{center}
\caption{\label{fig:ptch}$p^2_T$ distribution for $\kchrg$ candidate
events in the vacuum (left) and regenerator (right) beams. 
The data distribution is shown as dots, the $\kchrg(\gamma)$ 
signal MC (MC Sig) is shown as dotted histogram and the sum
of signal and background MC is shown as a solid histogram.}
\end{figure}

Figure \ref{fig:massch} shows
the $\pppm$ invariant mass distributions for the two beams; 
we require 488 MeV/$c^2 <$~m$_{\pppm} <$~508 MeV/$c^2$.  
Figure \ref{fig:ptch} shows the p$_T^2$ distributions; we require 
$p_T^2 <$~250 MeV$^2$/$c^2$.  The p$_T^2$ requirement is the only $\kchrg$
selection criterion that results in a systematic uncertainty in $\reepoe$.  We vary
the p$_T^2$ cut from 125 MeV$^2$/$c^2$ to 1000 MeV$^2$/$c^2$ and
assign a systematic uncertainty in $\reepoe$ of 0.10$\eu$ 
based on the change
in $\reepoe$.

To reduce our sensitivity
to details of the Monte Carlo simulation, 
we require track trajectories to be clear of a number of
physical apertures.  We require
that tracks point at least 2 mm into the CsI calorimeter away from the
edges of the Collar Anti detector that surrounds the beamholes
and at least 2.9 cm inside the
outer edge of the CsI calorimeter.  If the vertex position is upstream of the 
Mask Anti (MA, see Fig. \ref{fig:detector}), we require that the 
track position at the MA be less than 4 cm in $x$ and $y$
from the nominal beam center.  We cut away from wires at the edges of the 
drift chambers.  To reduce the 
possibility of $x$ and $y$ track candidate mismatches, we 
require that the projections of the tracks at the CsI calorimeter be separated by 
$6$ cm in $x$ and $3$ cm in $y$.  
We require that decays originate from within one of the beams
by requiring that the projection of the 
vertex ($x$,$y$) position along the kaon direction
reconstructs inside a 75 cm$^2$ square at the 
$z$ of the downstream edge of the regenerator.

We require
a minimum separation between the tracks in the $x$ and $y$ views at each 
drift chamber.  This cut is defined in terms of the DC cell through which the
track passes; we require that the tracks be separated by at least 3 cells
at each chamber.  
This track separation cut forms a limiting inner aperture and depends on the 
position of each wire within the drift chambers.  The wire spacing is known 
with an uncertainty of 20 $\mu$m.  There are variations in the actual wire spacing, which are 
measured in data, but are not simulated in the Monte Carlo.  
To determine the effect of
these variations, we convolve the track illumination with the wire-cell size to 
determine the number of events that migrate across the track separation cut in data 
but not in MC.  We find that the bias in $\reepoe$ is (-0.16 $\pm$ 0.12)$\eu$; the 
corresponding uncertainty in $\reepoe$ is $\pm$0.22$\eu$.

The effective regenerator edge, shown in 
Fig.~\ref{fig:regedge}, defines the upstream edge of 
acceptance for $\kchrg$ decays in the 
regenerator beam.  We find the effective regenerator edge by 
calculating the probability for two minimum ionizing pions 
to escape the last piece 
of scintillator without depositing enough energy to be vetoed.  
This calculation depends on the measured average energy deposit of a 
muon passing through the regenerator Pb module, the fraction of energy 
coming from the last 
piece of scintillator due to the geometry of the phototube placement 
on the Pb module,
the value of the trigger threshold, and the value of the offline cut on 
the energy deposit in the 
Pb module.
We find the effective regenerator edge to be (1.65 $\pm$ 0.4) mm upstream 
of the physical edge in 1997 and 
(0.7 $\pm$ 0.4) mm upstream of the physical edge in 1999.
The difference in
effective edges is due to different offline cuts on the energy deposit 
in the Pb module.  In 1997, the edge is defined by the trigger threshold.  
In 1999, a tight offline cut is applied.  We evaluate the uncertainty 
in this measurement by varying the trigger 
threshold and the fraction of energy coming from the last piece of 
scintillator by $\sim$15\% each.  
The 0.4 mm uncertainty in the position of the effective regenerator edge 
leads to an uncertainty in $\reepoe$ of 0.20$\eu$.

We estimate the systematic uncertainty in $\reepoe$ associated with the Monte
Carlo simulation of drift chamber efficiencies by generating separate 
sets of MC
in which scattering, DC efficiency maps, and accidental activity are 
turned off.  We take
10\% of the resulting variation in $\reepoe$, 0.15$\eu$,
to be the systematic uncertainty associated with the 
simulation of drift chamber efficiencies.  We vary the simulated drift 
chamber resolutions by 5\% and, from the resulting variation in $\reepoe$,
we assign a 
systematic error of 0.15$\eu$. 

The systematic uncertainties in $\reepoe$ associated with the $\kchrg$ analysis are
summarized in Table \ref{tb:chrgsyst}.  The total systematic uncertainty associated
with the $\kchrg$ analysis is 0.81$\eu$; this is reduced by $\sim 35\%$ from KTeV03.

\begin{table}[ht]
\centering
\begin{tabular}{l|c||cc|c}
\hline\hline
Source                   &\multicolumn{4}{c}{Error on $\reepoe$ ($\eu$)}    \\
                         & KTeV03 Result &\multicolumn{3}{c}{Current Result}  \\  
                         &             & 1997 & 1999               & Total  \\ \hline\hline
L1 and L2 Trigger        & 0.20 & 0.20 & 0.20               & 0.20          \\    
L3 Trigger               & 0.54 & 0.20 & 0.14               & 0.12          \\ \hline
Alignment and Calibration& 0.28 & \multicolumn{2}{c|}{0.20} & 0.20          \\
Momentum scale           & 0.16 & \multicolumn{2}{c|}{0.10} & 0.10          \\ \hline
$\ptsq$                  & 0.25 & \multicolumn{2}{c|}{0.10} & 0.10          \\ 
DC efficiency modeling   & 0.37 & \multicolumn{2}{c|}{0.15} & 0.15          \\ 
DC resolution modeling   & 0.15 & \multicolumn{2}{c|}{0.15} & 0.15          \\ \hline
Background               & 0.20 & \multicolumn{2}{c|}{0.20} & 0.20          \\ \hline
Wire Spacing             & 0.22 & \multicolumn{2}{c|}{0.22} & 0.22          \\ 
Reg Edge                 & 0.20 & 0.20 & 0.20               & 0.20          \\ \hline
Acceptance               & 0.79 & 0.87 & 0.25               & 0.41          \\
Upstream $z$             & ---  & 0.33 & 0.48               & 0.40          \\ \hline
Monte Carlo Statistics   & 0.41 & 0.28 & 0.28               & 0.20          \\ \hline
Total                    & 1.26 & 1.12 & 0.82               & 0.81          \\ \hline\hline    
\end{tabular}
\caption {Summary of systematic uncertainties in $\reepoe$ from the $\kchrg$ analysis. 
For errors which are evaluated individually for each year, the individual errors are 
listed in columns
and the total is the weighted average of the individual errors.
For those errors which are evaluated for the full dataset or taken to be the same for 
both years, only one number is listed. The
value of each systematic uncertainty from KTeV03 is provided for reference.}
\label{tb:chrgsyst}
\end{table}

\subsection{\label{sect:ana_neut}Neutral Reconstruction and Systematics}
To reconstruct $\kneut$ decays, we first identify four
clusters of energy in the calorimeter and reconstruct the energies and 
positions of the photons associated with each cluster.  A number of 
corrections are then made to the measured cluster energies based on our 
knowledge of the CsI calorimeter performance and the reconstruction algorithm.  
We use the cluster positions and energies along with the well-known pion mass 
to determine which pair of photons is associated with which neutral pion 
from the kaon decay and to calculate the decay vertex, the center-of-energy, 
and the $\pzpz$ invariant mass.  
The precision of the CsI calorimeter energy and position reconstruction is crucial to the
$\kneut$ analysis and has been improved significantly since KTeV03.
%We apply cuts to select $\kneut$ 
%decays, reduce background, and cut away from regions that are not well 
%simulated by the Monte Carlo.  
Section \ref{sect:csi} gives details of the CsI calorimeter reconstruction, Sec. 
\ref{sect:neutkine} describes the reconstruction of $\kneut$ decays,
Sec. \ref{sect:neutsel} describes the selection criteria for $\kneut$
decays, and
Sec. \ref{sect:esyst} describes the systematic uncertainties associated
with the CsI calorimeter energy reconstruction.

\subsubsection{\label{sect:csi}CsI Calorimeter Energy and Position Reconstruction}

The first step in reconstructing clusters is to determine the energy 
deposited in each crystal of the CsI calorimeter.  We convert the 
digitized information
to an energy using constants for each channel that are determined
from the electron calibration.  An in-situ laser, which delivers light at known 
intensities via quartz fibers to each CsI crystal, is used to calibrate the
DPMTs and to measure the less than 1\% spill-to-spill drifts in each channel's gain.  
The ``laser correction'' removes these spill-to-spill changes and is applied before any
clustering is performed.

%The laser correction, which is measured using an 
%in-situ laser and corrects for spill-to-spill drifts in each channel's gain, 
%is applied to each crystal energy.

We define a ``cluster'' as a 7$\times$7 array of small crystals or a 
3$\times$3 array of large crystals. Clusters near the boundary between
the small and large crystals (see Fig. \ref{fig:csilayout})
may contain both sizes of crystals; in this
case the cluster is defined as a $3\times3$ array of ``large'' crystals
where the energy deposit in four small crystals is summed to form a 
``large'' crystal as needed.   
Each cluster is centered on a ``seed crystal,'' 
containing the maximum energy deposit among the crystals in the cluster.
An initial approximation of the cluster energy is found by 
summing the energies of the crystals in the cluster.

The $x,y$ position of a cluster is reconstructed by calculating the fraction of 
energy in neighboring columns and rows of crystals in the cluster.  The 
$x,y$ position algorithm 
uses a map that is based on assuming a  uniform photon illumination across each 
crystal to convert 
these ratios to a position within the seed crystal.  The position maps
are made using isolated clusters from $\kneut$ data; no corrections
to the position are applied based on incident particle angle.
The final position is evaluated after all energy corrections are applied.

The raw cluster energy must be corrected for a number of geometric and 
detector effects.  We apply ``crystal-level'' corrections that adjust 
the energy in each crystal that makes up the cluster and
``cluster-level'' corrections, which are multiplicative corrections to the
total cluster energy.  Many of the crystal-level corrections rely on ``transverse
energy maps''; as a function of position within the seed crystal, these maps predict 
the distribution of energy among the crystals within a 
cluster .  They are made using isolated 
photon clusters from $\kneut$ data. The crystal-level and cluster-level energy corrections are
enumerated below.

\begin{enumerate}

\item \emph{Partial Clusters.} 
We correct for energy that is missing from the cluster because 
of crystal energies that are below the readout threshold or because 
portions of the 3$\times$3 
or 7$\times$7 cluster
are located in 
the beam holes or outside the calorimeter.  The energy in missing crystals
is estimated using the transverse energy maps.
The energy in crystals that were below threshold is estimated by a 
parameterization, which was determined from data, of the ratio of energy in a 
crystal to the readout threshold.
The fraction of the readout threshold energy predicted to be present in a
crystal decreases with distance from
the seed crystal and increases logarithmically with cluster energy.

\item \emph{Out of Cone.}
The $\sim$5\% ``out-of-cone correction'' is applied
because an electromagnetic shower is not fully contained by the 7$\times$7 small-crystal 
or 3$\times$3 large-crystal clusters.  
We determine the out-of-cone correction using the same GEANT 
simulation used to generate the Monte Carlo shower library (see Sec. 
\ref{sect:mcsim}).  The correction is parameterized by a quadratic function 
of the reconstructed distance of the cluster position from the center of 
the seed 
crystal and a linear function 
of the reconstructed energy.
The size of the correction varies by about 1\% across the face of a
crystal and by about 0.2\% per 100 GeV.  
There is no explicit dependence of the out-of-cone correction on incident
angle; because the reconstructed positions are not corrected for incident
particle angle, the angle effect is included implicitly in our 
parameterization as a function of reconstructed position.
The correction is generated separately for 
photons and electrons, and for small and large crystals.  
In KTeV03, the out-of-cone correction was determined for 
small and large crystals using 8 GeV GEANT showers, but there was 
no adjustment for the energy, position, or type of the incident particle.

\item \emph{Longitudinal Response.}
We correct the energy in each crystal for the $\sim$5\% non-uniformity 
of response along the length of each CsI crystal.
The longitudinal response of each CsI crystal is measured in ten 5-cm 
$z$ bins using cosmic ray muons that pass vertically through the CsI 
calorimeter.  These muons are detected by a cosmic ray hodoscope consisting
of three sets of 3 m-long, overlapping plastic scintillation counters placed
above and below the CsI calorimeter. 
Typically the crystal response increases with $z$ as the 
shower nears the PMT.  The measured CsI response is convolved with a
GEANT prediction of the shower's longitudinal distribution to 
correct the energy in each crystal.  The GEANT shower profiles 
are generated separately for photons and electrons.  There are individual 
profiles for each crystal position within the cluster; they are binned in 
local position relative to the center of the seed crystal and in the same
six logarithmic cluster energy bins used in the Monte Carlo (see Sec. \ref{sect:mcsim}).  
The mean shower depth for photons and electrons varies logarithmically with 
energy.  These crystal-by-crystal shower profiles are a significant improvement
to the longitudinal uniformity correction; in the KTeV03
analysis, the uniformity correction was applied at cluster level based only on a 
predicted average longitudinal energy distribution for a whole shower.

\item \emph{Shared Energy.}
For clusters that overlap, we must partition the 
energy in the shared crystals.  The ``overlap correction'' separates the 
energy deposited in two or more clusters that share crystals by using the
transverse energy maps to predict how much energy each particle contributed 
to the shared crystals.

The ``neighbor correction'' 
estimates the amount of underlying energy in each crystal that comes from 
nearby clusters that are less than 50 cm away but outside the 3$\times$3 
or 7$\times$7 cluster boundary.  The correction uses a 13$\times$13 map 
to predict the energy contribution from neighboring clusters.
This map is similar to the transverse energy maps but does not depend on position 
within the CsI crystal
and is generated using GEANT rather than data.

We correct clusters near the beam holes for extra energy that comes 
from nearby clusters that do not 
share crystals but which leak energy across the beam holes.  This correction
uses maps made using electrons from $\ke3$ data. 

\item \emph{Detector Effects.}
We correct for a number of detector effects including the 
observed transverse 
non-uniformity of energies across each crystal that remains after
the out-of-cone correction, the non-linearity of each 
channel with energy which remains after the longitudinal uniformity
correction, and global time variations in the CsI calorimeter response.
These corrections are measured using $E/p$ of electrons from $\ke3$ decays,
and are applied multiplicatively to the total cluster energy.
The ``transverse non-uniformity correction'' is made by dividing each cluster 
seed crystal into a 5$\times$5 grid and measuring the cluster energies of 
electrons in each of these position bins.  A multiplicative correction 
is applied to the total cluster energy based on the cluster's reconstructed 
position within the seed crystal.  The correction is normalized such that 
the average correction over each crystal (25 bins) is 1.0.  The 
``channel-by-channel linearity correction'' removes the residual energy 
non-linearity.  It is measured separately for each CsI calorimeter channel in data and 
Monte Carlo.  The ``spill-by-spill correction'' is applied to correct for 
time variations in the response of the calorimeter as a whole; it is
measured and applied separately for each spill.  Each of these corrections
has a maximum magnitude of less than 1\%.

\item \emph{Photon-Electron Differences.}
For the $\kneut$ analysis we apply a ``photon correction'' that
is designed to remove any residual differences between photons and the 
electrons that are used to calibrate the calorimeter.  The correction is 
based on photons from $\kneut$ and $\Kzzz$ decays.  It is measured separately 
for 1996, 1997, and 1999 in nine regions of the calorimeter by fitting each event for
the photon energies applying six (four) kinematic constraints for $\zzz$ ($\pzpz$).
The details of the kinematic fit are described in \cite{rsk3pi0}.
This correction is most important for photons with energies below 20 GeV;
its magnitude is less than 0.2\%. The photon correction is new for the current
analysis; no correction of photon-electron differences was applied in KTeV03.

\end{enumerate}

\begin{figure}
\begin{center}
\epsfig{file=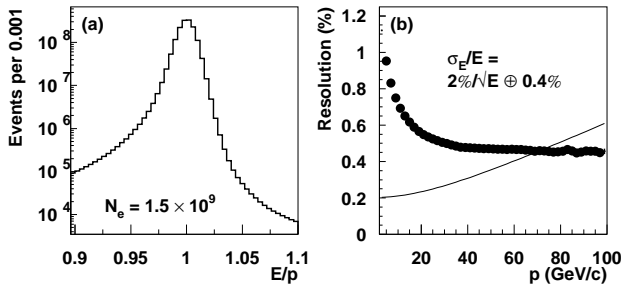,width=\linewidth}
\caption{$\Kethree$ electrons after all corrections. (a) $E/p$ for 1.5 
$\times$ 10$^9$ electrons. (b) Energy resolution.  The fine
curve shows the momentum resolution function that has been subtracted from 
the $E/p$ resolution to
find the energy resolution.}
\label{fig:resfinal}
\end{center}
\end{figure}

The quality of the calibration and the CsI calorimeter performance is evaluated by 
analyzing electrons from the $\ke3$ calibration sample with all 
corrections applied.  The electron calibration for 1996, 1997, and 1999 is 
based on 1.5 billion total electrons.  Figure \ref{fig:resfinal} shows the 
$E/p$ distribution and the energy resolution as a function of momentum 
of these electrons after 
all corrections.
The final energy resolution of the calorimeter is 
$\sigma_E/E \simeq 2\% / \sqrt{E} \oplus 0.4\%$, where $E$ is in GeV.

\subsubsection{\label{sect:neutkine}$\kneut$ Reconstruction}
$\kneut$ and $\Kzzz$ events are fully reconstructed using the positions 
and energies of the four or six photon clusters in the CsI calorimeter.  
The $\Kzzz$ reconstruction is almost identical to the $\kneut$ reconstruction,
but for simplicity this discussion will be in terms of the $\pzpz$ 
reconstruction.  Using cluster energies and positions, we are able to 
reconstruct the $z$ vertex of the kaon decay, 
the ($x$,$y$) components of the center-of-energy of the kaon, 
the kaon energy, and the $\pzpz$ invariant mass.

We must first determine which pairs of photons are associated  
in the $\kneut$ decay.  For four photons, there are three possible pairings.  
For each pairing we calculate $d_{12}$, the distance in $z$ between the 
$\piz$ decay vertex and  Z$_{CsI}$, the mean shower depth in the CsI crystals.
Using the
pion mass as a constraint, in the 
small angle approximation,
we find the distance for each pair of photons to be 
\beq
d_{12} \approx \frac{\sqrt{E_1E_2}}{m_{\pi^0}}r_{12},
\eeq
where
$r_{12}$ is the transverse distance
between the two photons at the CsI calorimeter.

For each pairing, we compare the calculated distance for each candidate pion.  
In most cases, 
only the correct pairing will give a consistent distance for both pions.  
The consistency of the measured distance is quantified using the 
pairing chi-squared variable:
\beq
%\chisqzz \equiv \sum_{i=1}^{N_{\pi^0}=2} \left( \frac{z_i - z_{avg}}{\sigma_i} \right)^2.
\chisqzz \equiv \left( \frac{d_{12} - d_{avg}}{\sigma_{12}} \right)^2 + \left( \frac{d_{34} - d_{avg}}{\sigma_{34}} \right)^2.
\label{eq:pairchisq}
\eeq
In Eq. \ref{eq:pairchisq}, $d_{ij}$ is the calculated distance for 
each pion, $d_{avg}$ is the weighted average of the distance $d_{ij}$ for both 
pions, and $\sigma_{ij}$ is the energy dependent vertex resolution for each pion.
We choose the pairing that gives the minimum value of $\chisqzz$.
Using Monte Carlo events, we find that this procedure selects the wrong
pairing for less than 0.01\% of $\kneut$ decays in the final event sample.
The $z$ vertex of the kaon decay is taken to be Z$_{CsI}$ - d$_{avg}$ for 
the best pairing.

We find the center-of-energy of the kaon decay at the CsI 
calorimeter plane
by weighting the position 
of each photon with its energy.  The $x$ and $y$ components of the 
center-of-energy are
\begin{equation}
       x_{\rm coe} \equiv \frac{ \sum x_i E_i }{ \sum E_i }~,
         ~~~~~~~~
       y_{\rm coe} \equiv \frac{ \sum y_i E_i }{ \sum E_i } ~,
        \label{eq:coe}
\end{equation}
where the sums are over all four photons.  The center-of-energy is the point 
at which the kaon would have intercepted the plane of the CsI calorimeter if it had not 
decayed, so we can calculate the ($x$,$y$) position of the decay vertex by 
assuming it lies on the line between the target and the center-of-energy.  
The $x$ coordinate of the kaon decay vertex is used to determine whether 
the kaon came from the regenerator or the vacuum beam.

The $\pzpz$ invariant mass is calculated from
the coordinates of the kaon decay vertex and the four photon positions and 
energies.  The kaon energy is calculated as the sum of the four photon energies.

\subsubsection{\label{sect:neutsel}$\kneut$ Selection}

The $\kneut$ event selection begins with three levels of trigger 
requirements during data taking.
The $\kneut$ Level 1 trigger requires that the total energy in the CsI calorimeter
be greater than 30 GeV.  The inefficiency in this trigger is studied using 
$\Kpmz$ decays from the $\kchrg$ trigger; the inefficiency at a given energy, 
$E_{total}$,
is the ratio of events with energy greater than $E_{total}$ for which the 
Level 1 trigger bit is not set
to the total number of events with energy greater than $E_{total}$. 
We find
inefficiencies ranging from (0.5-1.6)$\eu$ in the 40-160 GeV energy range used for the
analysis.  The impact of this inefficiency is slightly different in the vacuum and
regenerator beams because of small differences in the energy distributions. 
The resulting bias in $\reepoe$ is less than 0.02$\eu$, which we assign as a
systematic error.

The Level 2 trigger requirement is based on the ``hardware cluster counter''
(HCC)\cite{nim:hcc}.  The inefficiency in this trigger is measured using 
$\Kzzz$ decays
from a trigger that has no Level 2 requirement.  We reconstruct the $\Kzzz$
decays without any requirement on the HCC; the Level 2 inefficiency is the
ratio of the number of events that do not meet the HCC requirement
to the total number of events found in the offline reconstruction.  The bias
in $\reepoe$
produced by this inefficiency is determined using $\kneut$ MC.
The inefficiency is
simulated to within 10\% by the Monte Carlo, so we take 10\% of the 
measured bias as the systematic uncertainty in $\reepoe$.  The total 
uncertainty in $\reepoe$ associated
with the Level 2 trigger is 0.19$\eu$.

The inefficiency of the 
Level 3 $\kneut$ trigger is studied using ``random
accepts,'' a prescaled subset of the $\kneut$ trigger that has no Level 3
requirement.  We find no statistically signficant bias in $\reepoe$ and
quote an uncertainty of 0.07$\eu$ in $\reepoe$ based on the statistical precision of the
bias measurement.  

The offline selection criteria for the $\kneut$ sample are designed 
to select events that are cleanly reconstructed,
       to suppress background, and to select kinematic
and fiducial regions appropriate for the KTeV detector.
We evaluate the systematic uncertainty associated with each 
of the following requirements 
by loosening or removing the cut and evaluating the change in $\reepoe$.  

The energy of each CsI calorimeter cluster is required to be greater than 3 GeV because
the clustering corrections and MC simulation are not reliable at very
low energies.  
The minimum distance between the reconstructed positions of CsI 
calorimeter clusters
is required to be greater than $7.5$~cm
because it is difficult to separate the energy deposits in two very close clusters.
Clusters very near the beam holes are not as well 
reconstructed because of energy leakage across the beam holes and 
multiple overlapping or nearby clusters. In KTeV03, the inner CsI aperture was 
defined by the Collar Anti (CA) detector;
we now remove events with clusters having a seed crystal in the first ring of crystals 
around the beam holes. 
We do not find any systematic variation of $\reepoe$ with these requirements.

The variable describing the quality of the photon pairing, $\chisqzz$ 
(Eq. \ref{eq:pairchisq}),
is required to be less than 50.  This is a rather loose cut since 
more than 99\% of $\kneut$ events passing all other cuts have $\chisqzz$ 
values below 10.  The primary purpose of this cut is to reduce background 
from $\kppp$ events in which two of the photons escape the detector;
in this case it is likely that the missing
photons come from different pions causing the remaining photons to be
paired incorrectly.  The systematic uncertainty in $\reepoe$ associated
with this requirement is 0.14$\eu$.
  
The ``shape chi-squared'' variable, $\chisqshape$, is a measure of how well the
transverse energy distribution of each CsI calorimeter cluster matches the expected
distribution for a photon.  This variable, which is not a true chi-squared
because of correlations that are not considered, is calculated by comparing the
transverse energy distribution of each cluster to the transverse energy
maps described in Sec.~\ref{sect:csi}.  
The maximum value of $\chisqshape$ for each event 
is required to be less than 48.  The purpose of this cut is to remove background
from $\kppp$ events in which two or more photons overlap in the CsI calorimeter
and are reconstructed as a single cluster. In these cases the transverse distribution
of energy would tend to be different from that of a single photon cluster. 
The systematic uncertainty in $\reepoe$
associated with the shape chi-squared requirement is 0.15$\eu$.

We make a number of cuts on the veto detectors to 
reduce background.  We also use the calorimeter, spectrometer, and 
trigger hodoscope  as ``veto detectors'' by cutting on extra clusters,
tracks, and hits.
There is no systematic uncertainty associated with these requirements.

\begin{figure}
\begin{center}
\epsfig{file=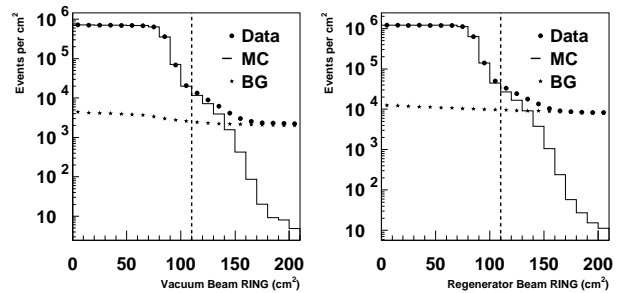,width=\linewidth}
\caption{$\kneut$ RING  distributions for data and signal MC in the
vacuum (left) and regenerator (right) beams.  The dashed line indicates 
our cut.}
\label{fig:ringcut}
\end{center}
\end{figure}

In the $\kchrg$ analysis we use p$_T^2$ to remove events in 
which the kaon scatters in the collimator or the regenerator.  This variable 
is not available for $\kneut$ decays since we do not measure the photon 
angles, so we use the ``ring number'' variable to reject scattered kaon 
decays.  Ring number is calculated using the 
center-of-energy of the reconstructed clusters, and is defined as
\beq
RING = 40000 \times Max(\Delta x_{coe}^{2} , \Delta y_{coe}^{2}),
\label{eqring}
\eeq
where $\Delta x_{coe}$ and $\Delta y_{coe}$ are the distances from the
center-of-energy to the center of the closest beam hole.
A change of $\Delta$RING$=1$ corresponds to an incremental area of 1 cm$^2$ 
centered on the beam hole.
Events with ring number less than 81 cm$^2$ should be from kaons decaying inside 
one of the two beams.  Figure \ref{fig:ringcut} shows the ring number 
distributions for both beams for data and Monte Carlo.  The ring number is 
required to be less than 110 cm$^2$; the systematic uncertainty in $\reepoe$
associated with this requirement is 0.27$\eu$.

The limiting apertures for $\kneut$ events are the CsI calorimeter inner
aperture at the beamholes, the CsI calorimeter outer aperture, the upstream edge 
in each beam,
and an effective inner aperture resulting from the 7.5 cm photon
separation requirement at the CsI calorimeter.  The CsI calorimeter inner and outer apertures
are defined by rejecting events in which a photon hits the innermost
or outermost ring of CsI crystals.  The upstream aperture in the vacuum
beam is defined by the Mask Anti and the upstream aperture in the regenerator
beam is defined by the lead module at the downstream edge of the regenerator.
The systematic errors associated with the precision of these apertures
are discussed in the KTeV03 paper\cite{prd03} and have not changed; the
individual values are listed in Table \ref{tb:neutsyst}.  The
total systematic uncertainty in $\reepoe$ associated with limiting apertures in
the $\kneut$ analysis is 0.48$\eu$. 

\begin{figure}
\begin{center}
\epsfig{file=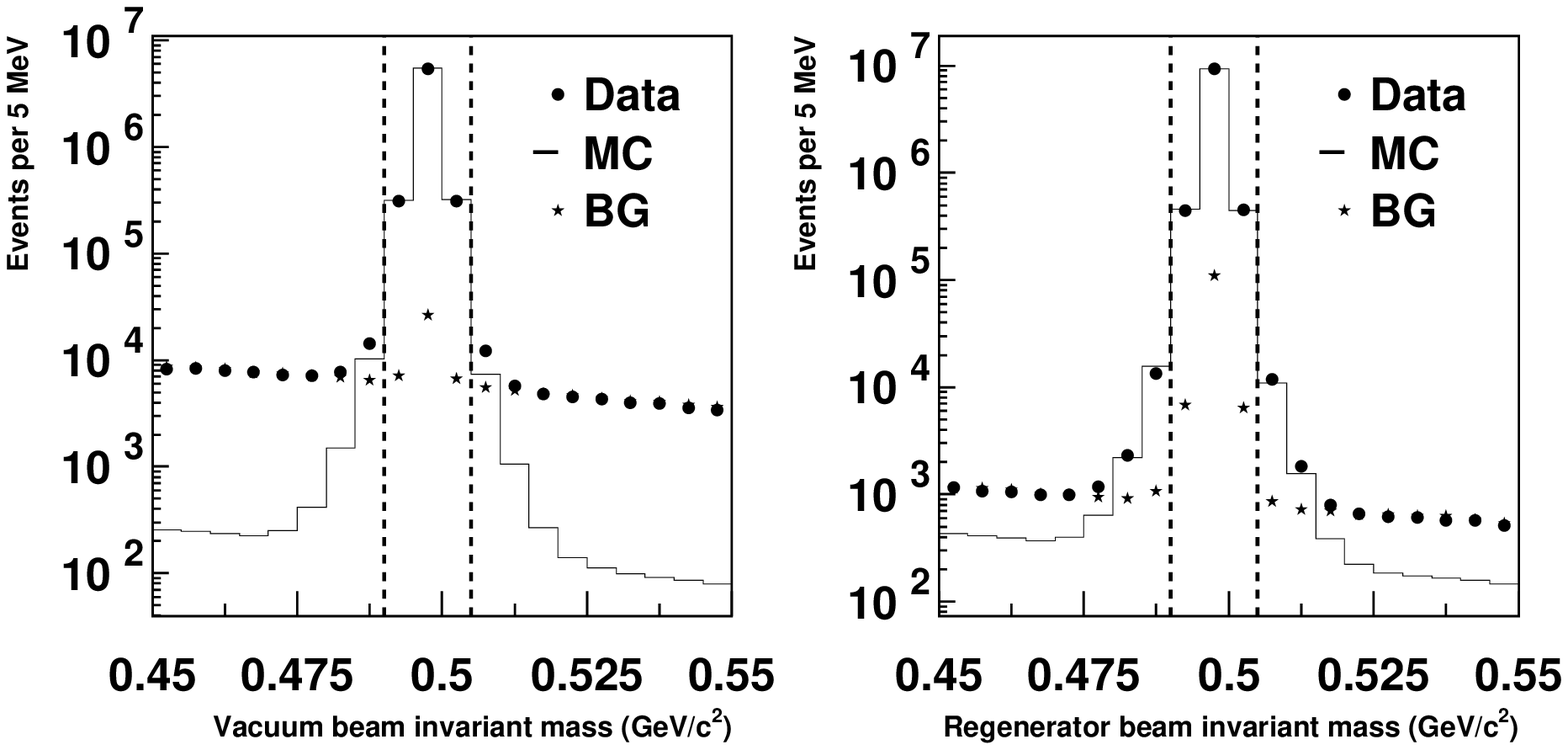,width=\linewidth}
\caption{$\kneut$ m$_{\pzpz}$ distributions for data and signal MC in the
vacuum (left) and regenerator (right) beams.  The dashed lines indicate 
our cuts.}
\label{fig:mkcut}
\end{center}
\end{figure}

Figure \ref{fig:mkcut} shows the reconstructed kaon mass distributions for both beams
for data and Monte Carlo.  The mass is required to be 
490 $\umass~<$ m$_{\pzpz}~<$~505~$\umass$.
The sidebands of the m$_{\pzpz}$ distribution are almost exclusively 
$\Kzzz$ background, with a small contribution from events in which the photons have been mispaired.
The peaking background at the kaon mass is from decays of kaons which scattered 
with non-zero angle in the regenerator and the defining collimators.  More details
on the background are given in Sect.~\ref{sect:bgsyst}.

\subsubsection{\label{sect:esyst}Energy Systematics}

The reconstruction of $\kneut$ decays depends entirely on the reconstruction
of energies and positions of photon showers in the CsI calorimeter.
Reconstructed quantities may depend upon the absolute energy scale or
the energy linearity of the CsI calorimeter.  We apply corrections that 
match the energy scale between data and Monte Carlo, 
and we assign
systematic uncertainties based on any disagreement in either absolute
energy scale or energy linearity between data and Monte Carlo.  The
procedures for matching the energy scale and evaluating the energy
systematics are described in this section. 

\begin{figure}
\begin{center}
\epsfig{file=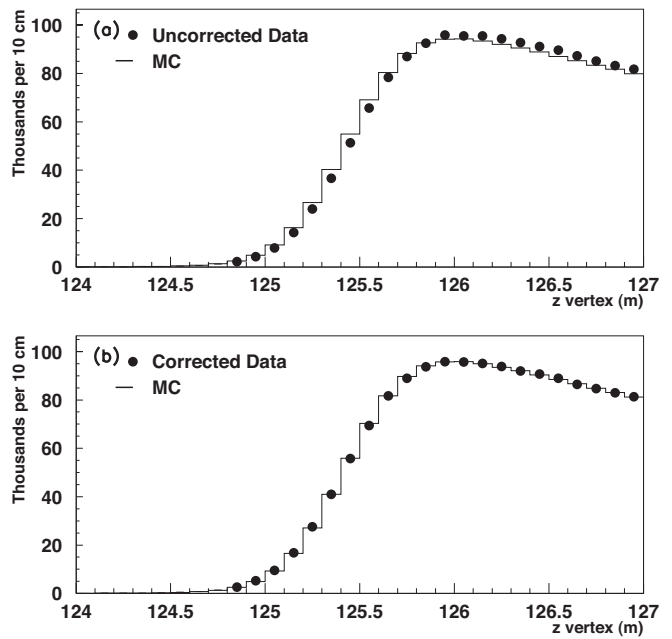,width=\linewidth}
\caption{Regenerator beam $\kneut$ $z$ vertex distribution near the
regenerator for 1999 data and Monte Carlo.  (a) Uncorrected data.  (b) Data
with energy scale correction applied.}
\label{fig:edgematch}
\end{center}
\end{figure}

The energy scale of the CsI calorimeter is set by the electron calibration,
but there is 
a small, residual difference in energy scale between data and Monte Carlo
for $\kneut$ events.
This difference is 
removed by adjusting the energy scale in data such that the sharp edge in 
the $z$ vertex distribution at the regenerator matches between data and 
Monte Carlo, as shown in Fig. \ref{fig:edgematch}.  
The correction is determined by sliding finely
binned $\kneut$ data and Monte Carlo $z$ vertex distributions in the 
regenerator beam past each other and using the Kolmogorov-Smirnov (KS) test
to determine how much the data must be adjusted to best match the MC.  
The correction is binned in kaon energy in the same 10 GeV energy
bins that are used to extract our results (see Sec. \ref{sect:fit}).
The same correction is applied to each cluster in an event.

\begin{figure}
\begin{center}
\epsfig{file=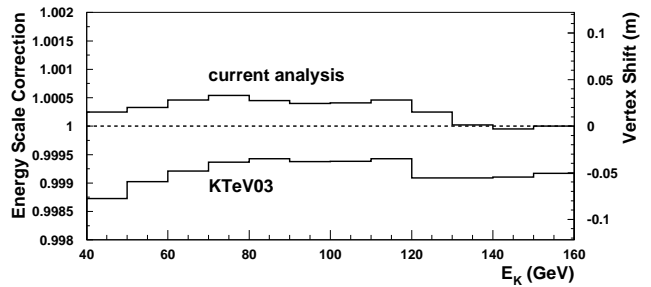,width=\linewidth}
\caption{Change in the final energy scale adjustment 
relative to KTeV03.  The dashed line represents no data-MC
mismatch. The $y$ axis on the right side of the plot shows
the data-MC $z$ vertex shift in meters.}
\label{fig:scalechange}
\end{center}
\end{figure}

The final energy scale adjustment is
shown as a function of kaon energy in Fig. \ref{fig:scalechange}.  
The average size of the $z$-vertex shift is $\sim$2.5 cm.
This corresponds to an average energy correction of $\sim$0.04\%, 
compared to $\sim$0.1\% in KTeV03.
As a result of improvements to the simulation and reconstruction of clusters, 
the required energy scale adjustment in the current analysis is smaller and less
dependent on kaon energy than in the KTeV03 analysis.

This final energy scale adjustment ensures that the energy scale matches 
between data and MC at the regenerator edge, but we must check whether the 
data and MC energy scales remain matched for the full length of the decay volume.  
Any non-linearity would result in different effective energy scales at different
decay points because of the correlation between the $z$ vertex and kaon energy
distributions.
We check the energy scale at the downstream end of the decay region by 
studying the $z$-vertex distribution of $\pzpz$ pairs produced
by hadronic interactions in the vacuum window and other downstream detector
elements in data and MC.  
%We call this sample ``vacuum window junk.''  
To verify that this type of 
production has a comparable energy scale to $\kneut$, we also study the 
$z$-vertex distribution of hadronic $\pzpz$ pairs produced 
in the regenerator. The $z$-vertex distribution of regenerator hadronic
events is Gaussian while the distribution of downstream events is more complicated,
as described below.
The methods for making the data-MC comparison in each case are described in the
following paragraphs.
%or ``regenerator junk.''

\begin{figure}
\begin{center}
\epsfig{file=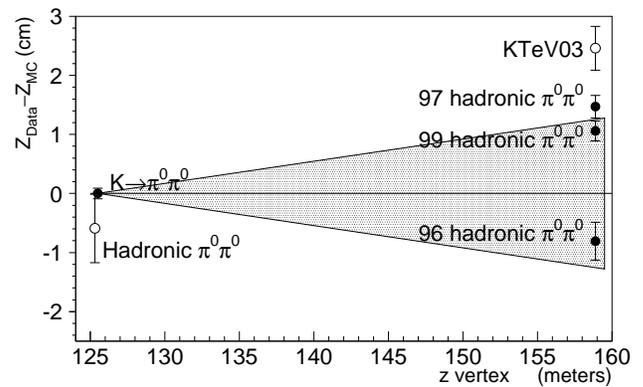,width=\linewidth}
\caption{Energy scale tests at the regenerator 
and vacuum window.  The difference between the reconstructed $z$ positions 
for data and MC is plotted for $\kneut$ events, and for hadronically produced $\pzpz$
pairs at the regenerator and the downstream detector elements.  The solid point at the 
regenerator edge is the $\kneut$ sample; there is no difference between data and MC by
construction.  The open point at the regenerator edge is the average shift 
of the hadronic regenerator samples for all three datasets.  The points at the 
vacuum window are the shifts for the downstream hadronic events for each dataset 
separately.  The shaded region shows the range of data-MC shifts covered 
by the total systematic uncertainty from the energy scale.
For reference, the data-MC shift at the vacuum window from KTeV03
 is also plotted.}
\label{fig:escalesyst}
\end{center}
\end{figure}

We compare the Gaussian $z$-vertex distributions of hadronically produced regenerator
events between data and MC by sliding the distributions past each other 
and using the chi-squared test.  The average data-MC difference is plotted in Fig. 
\ref{fig:escalesyst}; we find no significant data-MC mismatch in this 
sample.

\begin{figure}
\begin{center}
\epsfig{file=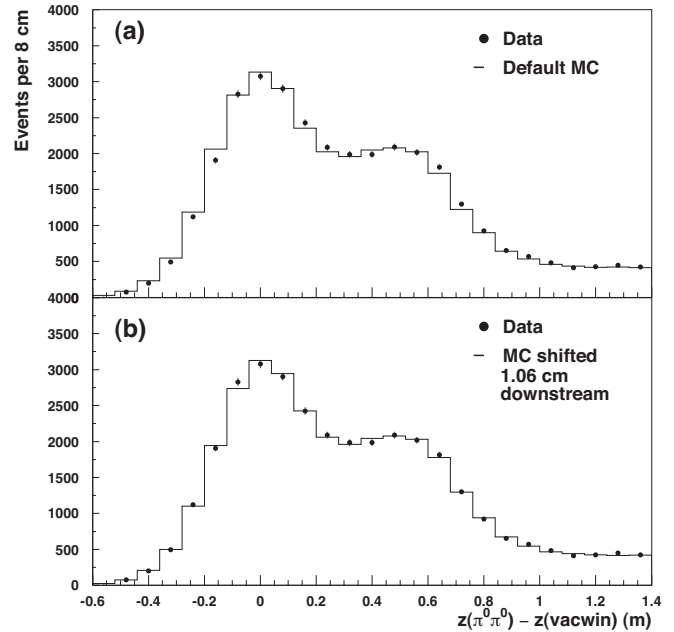,width=\linewidth}
\caption{$z$-vertex distributions 
of $\pzpz$ pairs produced hadronically in downstream detector elements for 1999 data 
and MC.  (a) Data (dots) and nominal MC (histogram).  
(b) Data (dots) and MC that is shifted 1.06 cm downstream to match 
the data (histogram).}
\label{fig:vacwinjunk}
\end{center}
\end{figure}

For the downstream hadronic events, we consider interactions in four separate detector 
volumes: the vacuum window, 
the upstream drift
chamber, and the two helium bags surrounding the drift chamber.  The production 
of $\pzpz$ pairs in each of these volumes is simulated separately; 
a fit is used to determine the relative contribution of each material,
and to find the difference between the data and MC $z$-vertex distributions.
The fit is performed separately for the 1996, 1997, and 1999 data samples.   
Figure \ref{fig:vacwinjunk}
shows the $z$-vertex distributions of downstream hadronic $\pzpz$ pairs
for 1999 data and MC, before and after 
the Monte Carlo data are shifted by the measured 1.06 cm data-MC difference.  
The $z$ shifts measured for 
each year are plotted in Fig. \ref{fig:escalesyst}.

To convert these shifts to an uncertainty in
$\reepoe$, we consider a
linearly varying energy scale distortion such that no adjustment 
is made at the regenerator edge and the $z$ shift at the vacuum window is 
that measured by the hadronic downstream sample.  This distortion
is shown by the shaded region in Fig. \ref{fig:escalesyst}.
We rule out energy scale distortions that vary non-linearly as a function 
of $z$ vertex because they introduce data-MC discrepancies in other distributions.
The systematic error on $\reepoe$ due to uncertainties in the $\kneut$ 
energy scale is 0.65$\eu$.

\begin{figure}
\begin{center}
\epsfig{file=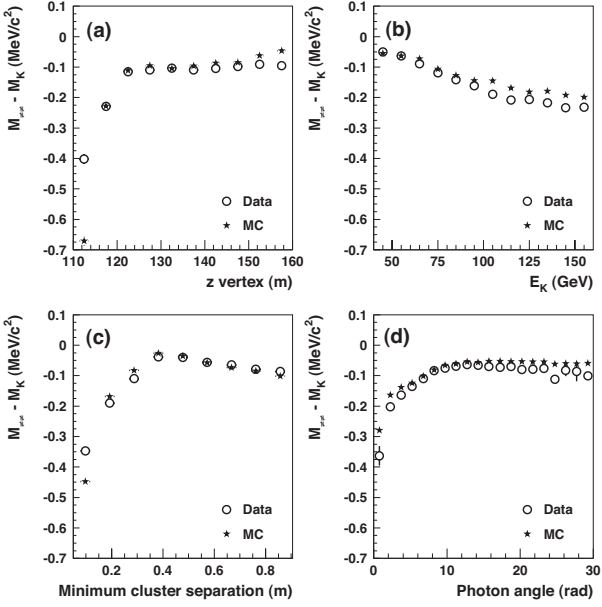,width=\linewidth}
\caption{Comparisons 
of the reconstructed kaon mass vs (a) $z$-vertex, (b) kaon energy, 
(c) minimum cluster separation, and (d) photon angle 
for 1999 data (circles) and MC (stars).
The values plotted are the difference between the reconstructed
kaon mass for each bin and the PDG kaon mass.}
\label{fig:nonlin1}
\end{center}
\end{figure}

\begin{figure}
\begin{center}
\epsfig{file=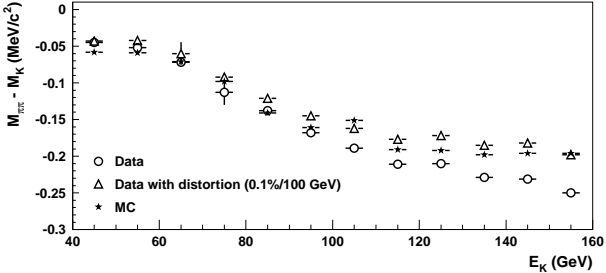,width=\linewidth}
\caption{Effect of 0.1\%/100 GeV distortion on M$_K$ vs E$_K$ for 1999 data.
The values plotted are the difference between the reconstructed
kaon mass for each bin and the PDG kaon mass.}
\label{fig:nonlin2}
\end{center}
\end{figure}

%Some reconstructed quantities in the analysis do not depend on the CsI
%calorimter energy scale, but are sensitive to energy non-linearities.
To evaluate the effect of energy non-linearities on the reconstruction,
we study the 
way the reconstructed kaon mass,
which does not depend on the absolute energy scale,
varies with reconstructed kaon energy, 
kaon $z$ vertex, minimum cluster separation, and incident photon angle.  
Data-MC comparisons for these distributions for the 1999 data sample 
are shown in Fig. 
\ref{fig:nonlin1}.  To measure any bias resulting from the nonlinearities
that cause the small data-MC
differences seen in these distributions, we investigate adjustments to the cluster energies 
that improve
the agreement between data and MC in the plot of reconstructed kaon mass 
vs kaon energy.  We find that a 0.1\%/100 GeV distortion produces the best data-MC 
agreement for the 1997 and 1999 datasets.  
Figure \ref{fig:nonlin2} shows the 
improvement in data-MC agreement with this distortion applied 
to 1999 data.
The 1996 dataset has slightly larger non-linearities; we find that a 
0.3\%/100~GeV distortion produces the best data-MC agreement for this dataset.
The data-MC agreement in the reconstructed kaon mass as a function of
kaon energy has been significantly improved compared to KTeV03,
where a 0.7\%/100 GeV distortion was required.

To evaluate the systematic error associated with these 
non-linearities, we apply the distortions to the data and find that $\reepoe$ 
changes by less than 0.2$\eu$ for all three datasets.
Properly weighting the three datasets, we 
find that the systematic uncertainty on $\reepoe$ due to energy 
non-linearities is 0.15$\eu$.

The systematic uncertainties in $\reepoe$ from the $\kneut$ analysis are
summarized in Table \ref{tb:neutsyst}.  The $\kneut$ analysis contributes
an uncertainty in $\reepoe$ of 1.55$\eu$, which is reduced by $\sim$23\% 
from KTeV03.

\begin{table}[ht]
\centering
\begin{tabular}{l|c||c|c|c|c}
\hline\hline
Source             &\multicolumn{5}{c}{Error on $\reepoe$ ($\eu$)}\\
                   &KTeV03 Result  &\multicolumn{4}{c}{Current Result}     \\
                   &             & 1996 & 1997 & 1999    & Total         \\ \hline\hline
L1 Trigger         & 0.10 & 0.01 & 0.01 & 0.03    & 0.02  \\ 
L2 Trigger         & 0.13 & 0.20 & 0.12 & 0.23    & 0.19  \\ 
L3 Trigger         & 0.08 & 0.20 & 0.04 & 0.05    & 0.07  \\ \hline
Ring Number        & 0.24 &\multicolumn{3}{c|}{0.27}& 0.27  \\ 
Pairing $\chi^{2}$ & 0.20 &\multicolumn{3}{c|}{0.14}  & 0.14       \\ 
Shape $\chi^{2}$   & 0.20 &\multicolumn{3}{c|}{0.15}  & 0.15       \\ \hline
Energy Nonlinearity& 0.66 & 0.10 & 0.10 & 0.20    & 0.15  \\ 
Energy Scale       & 1.27 & 0.45 & 0.82 & 0.59    & 0.65  \\ 
Position Reconstruction &0.35 & \multicolumn{3}{c|}{0.35} &  0.35  \\ \hline
Background         & 1.07 & 1.14 &\multicolumn{2}{c|}{1.06}&  1.07 \\ \hline
CsI Inner Aperture & 0.42 & \multicolumn{3}{c|}{0.42}     &   0.42 \\ 
MA Aperture        & 0.18 & \multicolumn{3}{c|}{0.18}     &   0.18 \\ 
Reg Edge           & 0.04 & \multicolumn{3}{c|}{0.04}     &   0.04 \\ 
CsI Size           & 0.15 & \multicolumn{3}{c|}{0.15}     &   0.15 \\ \hline
Acceptance         & 0.39 &\multicolumn{3}{c|}{0.48}& 0.48     \\ \hline
MC Statistics      & 0.40 & 0.75 & 0.37     & 0.41       &    0.25 \\ \hline\hline
Total              & 2.01 & 1.69 & 1.63     & 1.56       &    1.55 \\ \hline\hline    
\end{tabular}
\caption [Summary of systematic uncertainties in $\reepoe$ from the $\kneut$ analysis.]
{Summary of systematic uncertainties in $\reepoe$ from the $\kneut$ analysis.  For errors
which are evaluated individually for each year, the individual errors are listed in columns
and the total is the weighted average of the individual errors.
For those errors which are evaluated for the full dataset or taken to be the same for 
all years, only one number is listed. The value of each systematic uncertainty from KTeV03
is provided for reference.}
\label{tb:neutsyst}
\end{table}

\subsection{\label{sect:bgsyst}Background and Systematics}
Background to the $\Kpp$ signal modes is simulated using the Monte Carlo, 
normalized to data outside the signal region, and subtracted.  
There are two categories of background in this analysis: scattered $\Kpp$ events
and non-$\pi\pi$ background. We use decays 
from coherently regenerated kaons only; any kaons that scatter with non-zero angle
in the regenerator are treated as background.  This regenerator scattering 
background and background from kaons that scatter in the defining
collimators 
have the same momentum and  $p_T^2$ distributions 
for both $\kchrg$ and $\kneut$ decays.  
This background can be identified
using the reconstructed transverse momentum of the decay products in 
the charged decay mode. Therefore, the
scattering background is small in the charged mode, and we may use 
$\kchrg$ decays to tune the simulation of scattering background on which 
we must rely in the neutral mode.  

Non-$\pi\pi$ background is present because of misidentification 
of high branching-ratio decay modes.
The background to $\kchrg$ decays comes from $\ke3$ and $\km3$ decay modes.  
The background to $\kneut$ decays comes from $\Kzzz$ decays and hadronic 
interactions in the regenerator.  
The background estimation procedure and the
associated systematic uncertainties are described in detail in 
\cite{prd03}.

There is only one significant change to the background estimation
procedure since KTeV03.
Hadronic production of $K^*$ and $\Delta$ resonances
via $K_L+N\to K^*_S+X$ and $n+N\to \Delta+X$ are now included in
the $\kchrg$ regenerator beam background analysis; these background
sources were not considered in the KTeV03
analysis. The incident neutron spectrum
is assumed to be the same as that of the $\Lambda$ baryon, which
is measured in data. For $K^*_S$
decays, both $K^\pm\pi^\mp$ and $\pi^0 K_S, K_S\to\pi^+\pi^-$ modes
are simulated. The $K^*_S\to\pi^0 K_S$ background is normalized
using the transverse momentum
side band in the regenerator beam. The $K^*_S\to K^\pm \pi^\mp$
and $\Delta \to p^\pm \pi^\mp$ decays are normalized using mass sidebands
in the regenerator beam reconstructed assuming the vertex is located
at the regenerator edge. These two modes are seperated
using the momentum asymmetry distribution of the secondary particles.
The hadronic $K^*$ and $\Delta$ background samples have negligible 
contributions to
the signal region after all selection cuts, but including them
improves the description of mass and $p_T$ sidebands.

The background levels in $\kchrg$ are illustrated in Fig. \ref{fig:massch}
and Fig. \ref{fig:ptch}, and the background to $\kneut$ may be seen in Fig.
\ref{fig:ringcut} and Fig. \ref{fig:mkcut}.
Background contributes less than 0.1\% of
$\kchrg$ events and about 1\% of $\kneut$ events.  
Tables \ref{tb:chrgbgfrac} and \ref{tb:neutbgfrac} contain summaries of all the background 
fractions for each dataset.  There are some variations in background levels among the years due
to differences in trigger and veto requirements.
The systematic uncertainty
in $\reepoe$ due to background is 0.20$\eu$ from $\kchrg$ and 1.07$\eu$ from
$\kneut$.

%\begin{table}[ht]
%\centering
%\begin{tabular}{lcc}
%\hline\hline
%Source                 & Vacuum Beam & Regenerator Beam \\ \hline\hline
%\multicolumn{3}{l}{$\kchrg$}                           \\
%Regenerator Scattering & --          & 0.075\%          \\
%Collimator Scattering  & 0.008\%     & 0.008\%          \\
%$\ke3$                 & 0.032\%     & 0.001\%          \\
%$\km3$                 & 0.030\%     & 0.001\%          \\ 
%Total                  & 0.070\%     & 0.085\%          \\ \hline\hline
%\multicolumn{3}{l}{$\kneut$}                           \\
%Inelastic Scattering   & 0.128\%     & 0.175\%          \\
%Diffractive Scattering & 0.130\%     & 0.906\%          \\
%Collimator Scattering  & 0.120\%     & 0.091\%          \\
%$\Kzzz$                & 0.301\%     & 0.012\%          \\
%Photon Mispairing      & 0.008\%     & 0.007\%          \\
%Hadronic Production    & --          & 0.007\%          \\ 
%Total                  & 0.678\%     & 1.190\%          \\ \hline\hline
%\end{tabular}
%\caption{Summary of background levels for 1999.  Note that photon mispairing
%is not subtracted from the data and is not included in the total $\kneut$
%background sum}
%\label{tb:bgsum}
%\end{table}

\begin{table}[ht]
\centering
\begin{tabular}{l|cc|cc}\hline\hline
          & \multicolumn{2}{c|}{Vacuum Beam} & \multicolumn{2}{c}{Regenerator Beam} \\
Source                           & 1997    & 1999    & 1997    & 1999    \\ \hline \hline
Regenerator Scattering           & ---     & ---     & 0.073\% & 0.075\% \\
Collimator Scattering            & 0.009\% & 0.008\% & 0.009\% & 0.008\% \\
$\ke3$                           & 0.032\% & 0.032\% & 0.001\% & 0.001\% \\
$\km3$                           & 0.034\% & 0.030\% & 0.001\% & 0.001\% \\ \hline
Total Background                 & 0.074\% & 0.070\% & 0.083\% & 0.085\% \\ \hline\hline
\end{tabular}
\caption{Summary of $\kchrg$ background levels.}
\label{tb:chrgbgfrac}
\end{table}

\begin{table*}[ht]
\centering
\begin{tabular}{l|ccc|ccc}\hline\hline
          & \multicolumn{3}{c|}{Vacuum Beam} & \multicolumn{3}{c}{Regenerator Beam} \\
Source                             & 1996    & 1997    & 1999    & 1996   & 1997    & 1999    \\ \hline \hline
Regenerator Scattering & 0.288\% & 0.260\% & 0.258\% & 1.107\%& 1.092\% & 1.081\% \\
%Inelastic Scattering   & 0.153\% & 0.132\% & 0.128\% & 0.214\%& 0.186\% & 0.175\% \\
%Diffractive Scattering & 0.135\% & 0.128\% & 0.130\% & 0.893\%& 0.906\% & 0.906\% \\
Collimator Scattering  & 0.102\% & 0.122\% & 0.120\% & 0.081\%& 0.093\% & 0.091\% \\
$\Kzzz$                & 0.444\% & 0.220\% & 0.301\% & 0.015\%& 0.006\% & 0.012\% \\
Photon Mispairing      & 0.007\% & 0.007\% & 0.008\% & 0.007\%& 0.008\% & 0.007\% \\
Hadronic Production    & 0.002\% & 0.001\% & ---     & 0.007\%& 0.007\% & 0.007\% \\ \hline
Total Background       & 0.835\% & 0.603\% & 0.678\% & 1.209\%& 1.197\% & 1.190\% \\ \hline\hline
\end{tabular}
\caption{Summary of $\kneut$ background levels.  Note that
photon mispairing is not subtracted from the data and is not included in the total background sum.}
\label{tb:neutbgfrac}
\end{table*}

\subsection{Data Summary}

The numbers of 
events collected in each beam are summarized in Table \ref{tb:eventtot}.
After all event selection requirements are applied and  
background is subtracted,  
we have a total of 25 million vacuum beam $\kchrg$ decays and 
6 million vacuum beam $\kneut$ decays.

\begin{table}[ht]
\centering
\begin{tabular}{l|cc}
\hline\hline
             & Vacuum Beam & Regenerator Beam \\ \hline
$\kchrg$     & 25107242    & 43674208         \\
$\kneut$     & 5968198     & 10180175         \\ \hline\hline
\end{tabular}
\caption {Summary of event totals after all selection criteria and background
subtraction.}
\label{tb:eventtot}
\end{table}

\section{\label{sect:accandfit}Acceptance and Fitting}
\subsection{Acceptance Correction and Systematics}

We use the Monte Carlo simulation to estimate
the acceptance of the detector in
momentum and $z$-vertex bins in each beam.  We evaluate the quality of
this simulation by comparing $z$-vertex distributions
in the vacuum beam between data and Monte Carlo.  To account for
small differences in the energy spectrum between data and Monte Carlo, we
reweight the distributions,
using the same 10 GeV/$c$ momentum
bins used by the fitter (see Section \ref{sect:fit}), by
adjusting the number of MC events in each bin so that the data and MC
kaon momentum distributions agree.  
We fit the data-MC ratio of $z$-vertex distributions to a line, and
call the slope of this line, $s$, the acceptance ``z-slope.'' We use this
z-slope to evaluate the systematic error on $\reepoe$.

A z-slope affects the value of $\reepoe$ by producing a bias between the 
regenerator and vacuum beams because of the different $z$ vertex 
distributions in the two beams.  A good approximation of the bias on 
$\reepoe$ is $s\Delta z$/6 where $\Delta z$ is the difference of the mean 
$z$ values for the vacuum and regenerator beam $z$ vertex distributions.  
The factor of 6 converts the bias on the vacuum-regenerator beam ratio to 
a bias on $\reepoe$.  The values of $\Delta z$ are 5.6 m for the $\kchrg$ sample and 
7.2 m for the $\kneut$ sample.  We use the measured bias 
on $\reepoe$ and the statistical error on that measurement to assign
a systematic uncertainty in $\reepoe$.

\begin{figure*}
\begin{center}
\epsfig{file=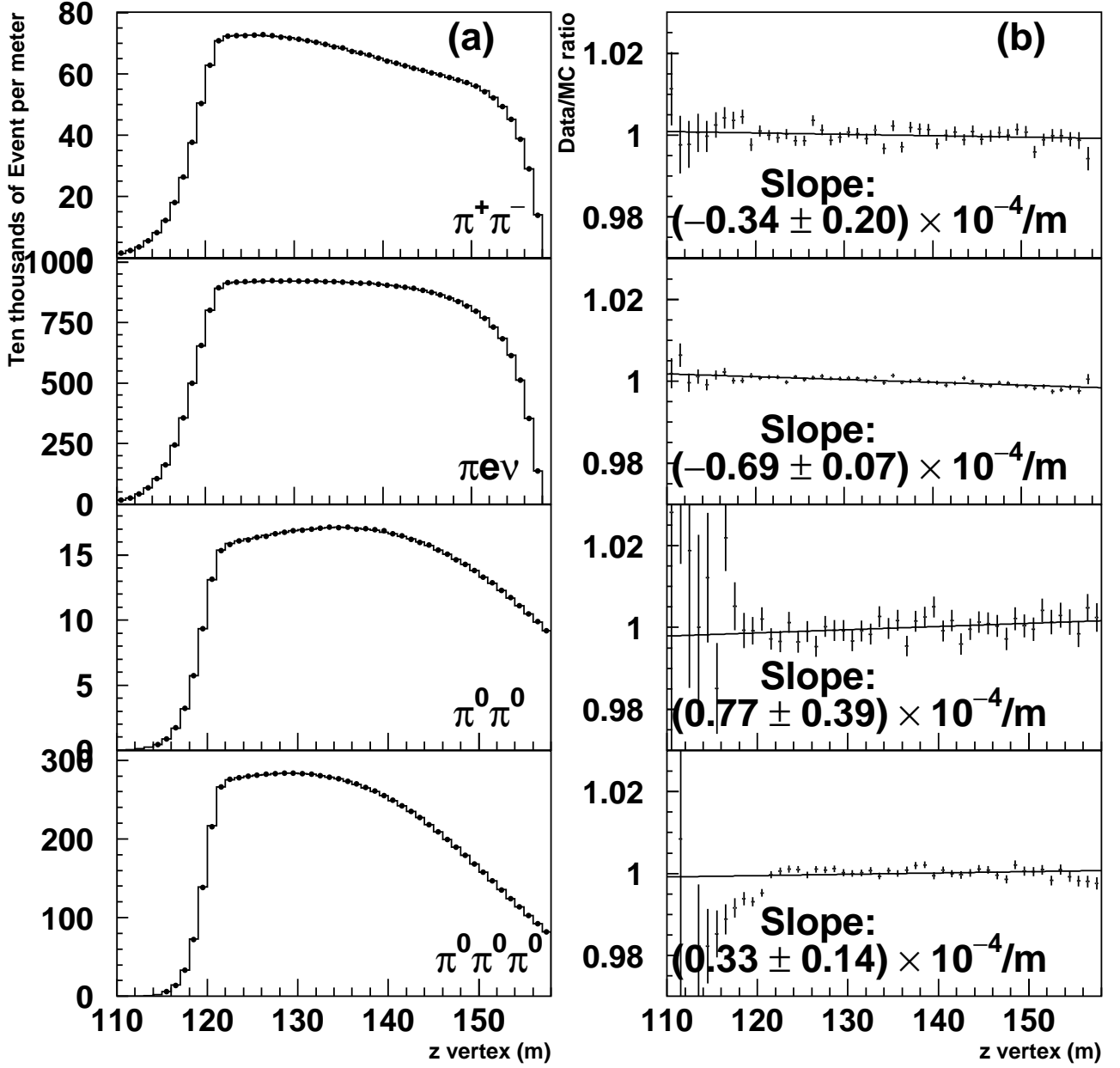,width=\linewidth}
\caption{(a) Comparison of the vacuum beam $z$ distributions for
data (dots) and MC (histogram).  (b) The data-to-MC ratios
are fit to a line, and the z-slopes (see text) are
shown.  All distributions are for the full data sample used in this
analysis.}
\label{fig:zslopes}
\end{center}
\end{figure*}

The z-slopes for the full dataset are shown in Fig. \ref{fig:zslopes}.
We use the 25 million vacuum beam $\kchrg$ decays to measure the z-slope 
in the charged decay mode.  The
uncertainty in $\reepoe$ associated with this z-slope is 0.41 $\eu$.  
We assign an additional uncertainty of 0.40 $\eu$ based on a $\reepoe$ 
fit which excludes $\kchrg$ decays from the region upstream of the MA.
This region is very sensitive to  the value of the MA apperture cut
(see Sec.~\ref{sect:ana_chrg}), and, since it lies upstream of the regenerator
edge, there are no regenerator beam decays to compensate for
this dependence.  Combining these two uncertainties, the systematic
uncertainty in $\reepoe$ associated with the acceptance correction is 0.57$\eu$.

We also measure
the data-MC $z$-slope in the high statistics $\ke3$
decay mode and find a slope that is similar in magnitude to the systematic
uncertainty from $\kchrg$. 
We do not use the $\ke3$ z-slope to set the systematic error
because it is sensitive to different detector effects and has different particle types 
in the final state than $\kchrg$.

We use 88 million $\Kzzz$ decays to measure the z-slope 
in the neutral decay mode.  This mode has the same type of particles in the 
final state as $\kneut$, and it is more sensitive than $\pzpz$ to potential 
problems in the reconstruction due to close clusters, energy leakage at 
the CsI calorimeter edges, and low photon energies. We assign an uncertainty on $\reepoe$ 
from the neutral mode acceptance
of 0.48 $\eu$ based on the $\Kzzz$ z-slope. We also measure the z-slope in 
$\kneut$ 
decays and find that the results are consistent with those from $\Kzzz$ 
decays.  There is no significant change in $\reepoe$ when the region upstream 
of the MA is excluded in the neutral mode, so no additional systematic uncertainty 
is required.

\subsection{\label{sect:fit}Fitting and Systematics}
The value of $\reepoe$ and other kaon parameters
$\delm$, $\tauS$, $\phiep$, and $\imepoe$ are determined
using a fitting program. The fitting procedure
is to minimize $\chi^2$ between background subtracted data
and a prediction function. The prediction
function uses the detector acceptance determined with the Monte Carlo
simulation. The fits are  performed in $10$~GeV/$c$ kaon momentum
bins. There is no $z$ binning to determine $\reepoe$, while 
a $z$-binned fit is performed
to measure the other kaon parameters.
Uncertainties from the fitting procedure are mainly related to regenerator properties and 
the dependence of the result on external parameters.

Neglecting the  contribution from  $\KS$ produced
at the target (called target-$\KS$),
the number of $K\to\pi\pi$
events in the vacuum beam for a given $p,z$ is
\begin{equation} \label{eq:vac}
 N^{\pi\pi}(p,z) \sim {\cal F}(p) |\eta|^2 
\exp \left(-\frac{\textstyle t}{\textstyle \tau_L}\right),
\end{equation}  
where $t= (z-z_{reg}) m_K/p $ is the measured proper time relative
to decays at the regenerator edge, $\eta = \eta_{+-} = \epsilon + \epsilon'$
($\eta = \eta_{00} = \epsilon - 2\epsilon'$) for charged (neutral) decays and
${\cal F}(p)$ is the kaon flux. The fitting program includes the contribution of 
target-$\KS$ by using a phenomenological model for $\Kz$/$\Kzbar$ production at 
the
target and propagating the kaon states up to the 
decay volume.
The model of target-$\KS$ production is checked by floating the 
$\Kz$/$\Kzbar$ flux ratio 
in the fit.
The fitted fraction of target-$\KS$ deviates from the model  
by (2.5 $\pm$ 1.6)\%.
The associated systematic uncertainty 
in $\reepoe$ is $\pm$0.12$\eu$.  

The number of events in the regenerator beam is
\begin{equation} \label{eq:reg}
\begin{array}{l} 
N^{\pi\pi}(p,z) \sim {\cal F}(p) T_{reg}(p)  \\ 
 \times  \left[
  |\rho(p)|^2\exp\left( -\frac{\textstyle t}{\textstyle \tau_S} \right)
+ |\eta|^2 \exp\left( -\frac{\textstyle t}{\textstyle \tau_L} \right) \right.\\
+ \left. 2 |\rho(p)||\eta|\cos\left(\delm t + \phi_\rho(p) - \phi_\eta \right)
   \exp\left( -\frac{\textstyle t}{\textstyle \tau_{ave}} \right) \right], \\
\end{array}
\end{equation}
where  $\rho(p)$ is the momentum-dependent 
coherent regeneration amplitude, $\phi_\rho(p) = \arg(\rho)$, 
$1/\tau_{ave} = (1/\tau_S + 1/\tau_L)/2$ and $T_{reg}(p)$ is the relative kaon
flux transmission in the regenerator beam. The prediction function 
accounts for decays inside the regenerator by using the effective
regenerator edge (Fig.~\ref{fig:regedge}b) as the start of the decay region.

The parameters from Eqs.~\ref{eq:vac},\ref{eq:reg} are determined as discussed
below. The kaon flux, ${\cal F}(p)$, is a free parameter for each of the twelve
$10$~GeV/$c$
momentum bins. Separate kaon fluxes are allowed for charged and neutral
decays to account for slight differences in the data samples, so there
are a total of $12\times 2=24$ free fit parameters to describe the kaon flux. 
The flux ratio  of decays in the vacuum and regenerator beams is, however, the same   
in both charged and neutral decay modes.  
The 1996 $\kneut$ data has no corresponding $\kchrg$ data, so it is possible that 
there could be small
differences in the flux ratio between the two years which do not cancel in the fit.  
We assign an 
uncertainty in $\reepoe$ of $\pm$0.03$\eu$ from this possibility, following~\cite{prl:pss}.  

\begin{figure}
\begin{center}
\epsfig{file=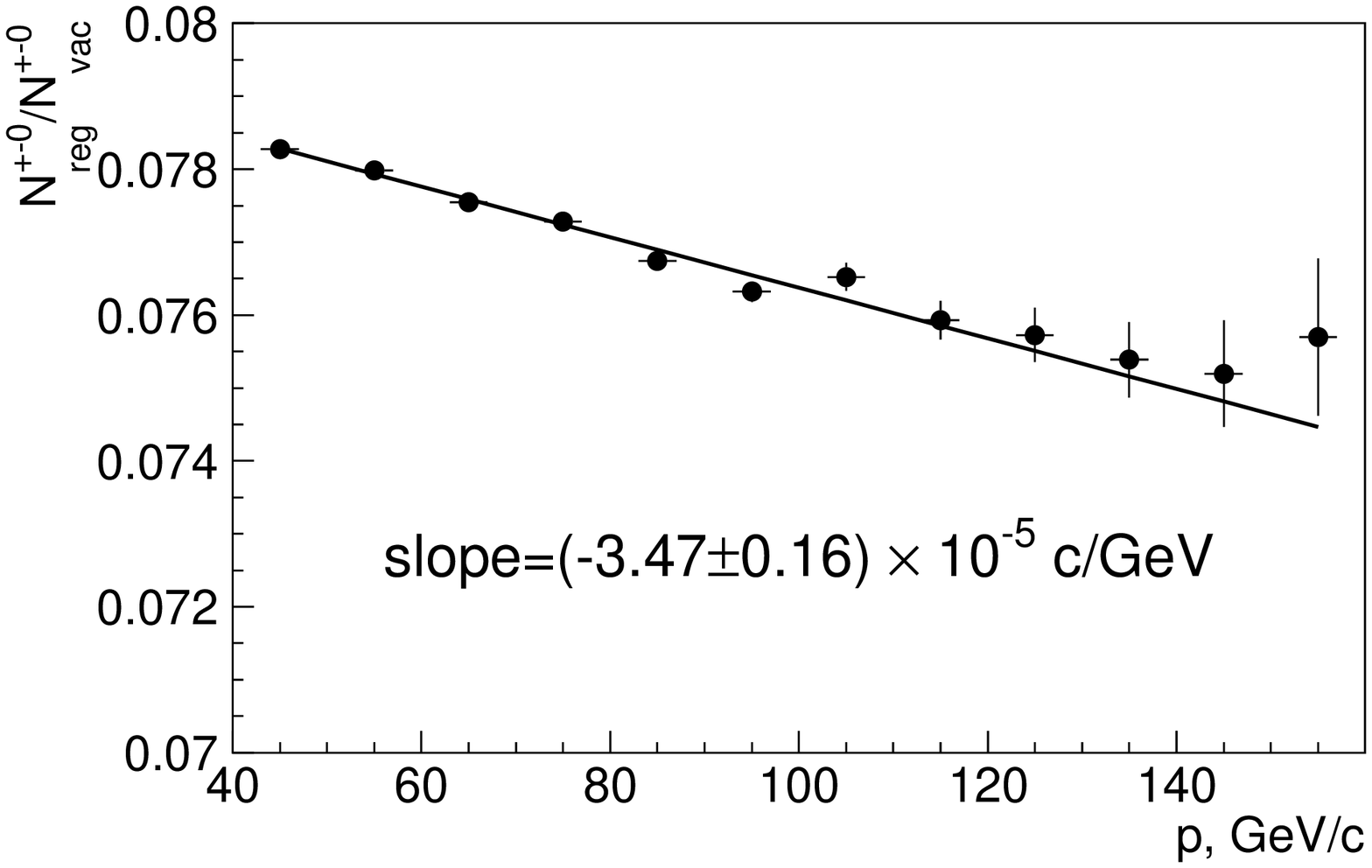,width=\linewidth}
\end{center}
\caption{\label{fig:regtrans}Ratio of $K_L\to \pi^+\pi^-\pi^0$ decay rates
in the regenerator to vacuum beam as a function of a kaon momentum. The uncertainty
on the slope is statistical uncertainty of the measurement.}
\end{figure}

The relative kaon flux attenuation in the regenerator beam, $T_{reg}$, results from
the shadow absorber and the regenerator itself. 
The attenuation is measured
directly from data by comparing the rate of $K_L \to \pi^+\pi^-\pi^0$ decays
in the vacuum $(N^{+-0}_{vac})$ and regenerator $(N^{+-0}_{reg})$ beams. 
As mentioned earlier, a dedicated trigger was introduced
in 1999 to improve the statistical precision of this measurement;
the improved measurement is applicable for all years of data taking.
The attenuation is found to be a linear function of $p$ for a
kaon momentum range of $40-160$~GeV/$c$ as shown in Fig.~\ref{fig:regtrans}. 
For the 61.5 GeV/$c$ average momentum of $K_L\to \pi^+\pi^-\pi^0$ decays,
the transmission
is $T_0 = (7.771\pm 0.004)\%$.  The slope of the momentum dependence is 
$\alpha_T = (-3.5 \pm 0.2 )\times 10^{-5}$~$c$/GeV 
where the errors represent total experimental uncertainties.
The uncertainty on the momentum dependence 
of the regenerator attenuation corresponds to a 0.08$\eu$ uncertainty 
in $\reepoe$.

The regeneration amplitude $\rho$ is related to the difference between the 
forward kaon-nucleon
scattering amplitudes for $\Kz$, $f(0)$,  
and $\Kzbar$, $\bar{f}(0)$,~\cite{prd:731}: 
\begin{equation} \label{eq:fminus}
f_- = \hbar \frac{\textstyle f(0) - \bar{f}(0)}{\textstyle p}.
\end{equation}
The KTeV regenerator is composed mostly of plastic scintillator with two 
thin lead plates in the last module (see Fig~\ref{fig:regedge}a).  
The dominant source for regeneration is forward scattering in carbon with 
small additional contributions from hydrogen and lead. The regeneration 
in hydrogen and lead is fixed 
in the fit while 
the parameters describing carbon regeneration are allowed to float.

For an isoscalar target ($C_{12}$) and high kaon momentum, $f_-$ can be 
approximated by a single Regge trajectory~\cite{pr:gilman}
and the magnitude of $f_-$ varies  with kaon  momentum as a power law,  
$|f_-(p)| \sim p^\alpha$. The analyticity of the forward scattering 
amplitude relates the magnitude of $f_-(p)$ and its phase, $\arg(f_-)$. 
The magnitude of $|f_-|$ at $p=70$~GeV/$c$ and the power law $\alpha$ are 
the two free parameters describing regeneration in the fit.
We estimate the systematic error from the analyticity 
assumption by allowing $\phi_{\rho}$ to deviate 0.25$\degs$ from analyticity; 
the associated uncertainty in 
$\reepoe$ is 0.07$\eu$.  

For scattering off complex nuclei, the effects of  
nuclear screening corrections are important
and the regeneration amplitude cannot be described by    
a single power law. 
We vary the screening models in the fit and find that 
the associated $\reepoe$ uncertainty is
0.24$\eu$. 
More details on the screening corrections and the analyticity relation for 
the scattering amplitude are given in Appendix~\ref{sec:regeneration}.

The $K_L$ lifetime, $\tau_L$, is taken from~\cite{pdg06};
the uncertainty in $\reepoe$ due to the uncertainty in 
this measurement is 0.01$\eu$.
The values of $\delm$ and $\tauS $ are fixed to our measurements 
(Eq.~\ref{dmts:sw}) for the $\reepoe$ fit and are floated in the $z$-binned
fit.
The uncertainty in $\reepoe$ due to the values of $\delm$ and $\tauS$ 
used in the fit is 0.11$\eu$.  

The systematic uncertainties in $\reepoe$ associated with fitting are summarized 
in Table \ref{tb:fitsyst}.
The total systematic uncertainty in $\reepoe$ from fitting is 0.31$\eu$.

\begin{table}[ht]
\centering
\begin{tabular}{|l|c|} \hline
Source         & Error on $\reepoe$    \\
               & ($\eu$)               \\ \hline
Regenerator transmission & 0.08        \\
Target-$\KS$             & 0.12        \\
$\delm$ and $\tauS$      & 0.11        \\
Regenerator screening    & 0.24        \\
$\phi_{\rho}$ (analyticity) & 0.07      \\
1996 $\KS$/$\KL$ flux ratio & 0.03     \\
$\tau_L$                 &  0.01       \\ \hline
Total                    &  0.31       \\ \hline
\end{tabular}
\caption{Summary of systematic uncertainties in $\reepoe$ associated with fitting.}
\label{tb:fitsyst}
\end{table}

% DISCUSS RHO. GIVE A FIGURE FOR SCREENING CORRECTIONS for |RHO| and ARG|RHO|

% brief, explain what fitter does and the different fits but not details of fitter

\section{\label{sect:results}Results}

\subsection{Measurement of $\reepoe$}
In the KTeV fit for $\reepoe$, the inputs are the observed number of 
$\kchrg$ and $\kneut$ decays in each of twelve 10 GeV/$c$ momentum bins.  
The kaon fluxes for
$\kchrg$ and $\kneut$ in each momentum bin, the regeneration parameters,
and $\reepoe$ are free parameters.  CPT symmetry is assumed by setting the
phases $\phipm$ and $\phizz$ equal to the superweak phase.
The final KTeV result is:
\bqa
\reepoe & = & [19.2 \pm 1.1(stat) \pm 1.8(syst)]\eu \nonumber \\
        & = & [19.2 \pm 2.1]\eu. 
\eqa
The fit $\chi^2$ is $\chi^2/\nu = 22.9/21$.
The systematic uncertainties in $\reepoe$ are summarized in 
Table \ref{tb:systsummary}. This result corresponds to a
particle-antiparticle partial decay rate asymmetry of
\begin{equation}
\frac{\Gamma(\Kz \to \pi^+\pi^-) - \Gamma(\Kzbar \to \pi^+\pi^-) }
{\Gamma(\Kz \to \pi^+\pi^-) - \Gamma(\Kzbar \to \pi^+\pi^-)}
= (6.2 \pm 0.6) \times 10^{-6}. 
\end{equation}

\begin{table}[ht]
\centering
\begin{tabular}{l|cc} 
\hline\hline
Source                & \multicolumn{2}{c}{Error on $\reepoe$ ($\eu$)} \\
                      & $\kchrg$ & $\kneut$                             \\ \hline
Trigger               & 0.23     & 0.20                                 \\
CsI cluster reconstruction & --- & 0.75                                 \\
Track reconstruction  & 0.22     & ---                                  \\
Selection efficiency  & 0.23     & 0.34                                 \\
Apertures             & 0.30     & 0.48                                 \\
Acceptance            & 0.57     & 0.48                                 \\
Backgrounds           & 0.20     & 1.07                                 \\
MC statistics         & 0.20     & 0.25                                 \\ \hline
Total                 & 0.81     & 1.55                                 \\ \hline
Fitting               & \multicolumn{2}{c}{0.31}                       \\ \hline
Total                 & \multicolumn{2}{c}{1.78}                          \\ \hline\hline
\end{tabular}
\caption{Summary of systematic uncertainties in $\reepoe$.  
See Tables \ref{tb:chrgsyst} and \ref{tb:neutsyst}
for more details on the errors from the $\kchrg$ and $\kneut$ analyses, 
respectively.}
\label{tb:systsummary}
\end{table}

\begin{figure}
\begin{center}
\epsfig{file=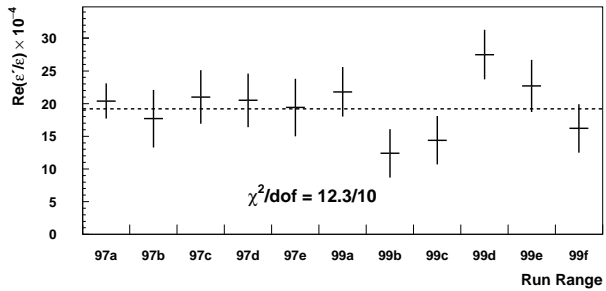,width=\linewidth}
\caption{$\reepoe$ in subsets of the data sample.  All points are
statistically independent.  The dashed line indicates the value of $\reepoe$ for the full data sample.  The 97a 
run range includes the 1996 $\kneut$ data.}
\label{fig:eperuns}
\end{center}
\end{figure}

We perform several checks of our result by dividing the data into subsets and 
checking the
consistency of the $\reepoe$ result in the various subsets.  To check for 
time dependence, we break the data into eleven run ranges with roughly 
equal 
statistics.  There are five run ranges in 1997 and six in 1999.  Since the 
1996 $\kneut$ data does not have any corresponding $\kchrg$ data, we combine
it with the neutral mode data in the first 1997 run range.  As shown in 
Fig. \ref{fig:eperuns},  we find 
consistent results in all of the run ranges.  The combined result for 1996
and 1997 data is $\reepoe = [20.0 \pm 1.7(stat)] \eu$, which can be compared 
directly to the KTeV03 result of 
$\reepoe =  [20.7 \pm 1.5(stat)] \eu$ \cite{prd03}.  
The decrease in statistical precision is due primarily to
the removal of $\kneut$ events with a cluster near one of the beam holes (see
Sect. \ref{sect:neutsel}).

\begin{figure}
\begin{center}
\epsfig{file=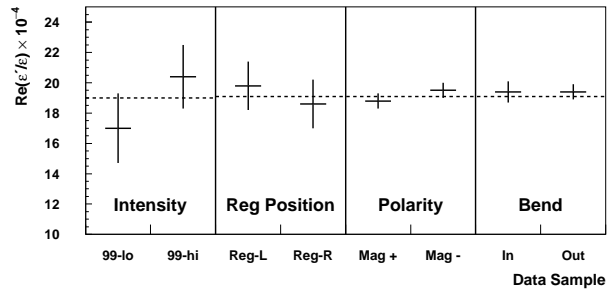,width=\linewidth}
\caption{$\reepoe$ consistency with beam intensity, regenerator position, 
magnet polarity, and track bend.   
The intensity subsets are for the 1999 data only.
Reg-left and reg-right refer to the position of the regenerator beam in 
the detector.  These subsets are for the full data sample.  Mag$+$ and Mag$-$ are
the magnet polarity and in/out are the bend of the two tracks in the magnet.  
In each of these subsets the $\kneut$ sample is common to both fits; 
the errors are estimated by taking the quadrature difference with
the error for the full dataset.  The dashed lines
indicate the value of $\reepoe$ in the appropriate full data sample.}
\label{fig:epechecks}
\end{center}
\end{figure}

In 1999 we took data at high and low proton beam intensity so we are able to check 
for dependence of $\reepoe$ on beam intensity.
About 43\% of the 1999 data were collected at low intensity (defined as 
less than $1.25 \times 10^{11}$ protons/s) while 57\% were collected at 
high intensity (greater than $1.25 \times 10^{11}$ protons/s).  
The average proton rate in the low intensity sample is 
$\sim 1 \times 10^{11}$ protons/s and the average rate at high intensity 
is $\sim 1.6 \times 10^{11}$ protons/s.  
The average rate for 1996 and 1997 data is $\sim 1.5 \times 10^{11}$
protons/s.
Figure \ref{fig:epechecks} shows the 1999 $\reepoe$ result for
low and high intensities; there is no evidence for intensity dependence
of the result.

%\begin{figure}
%\begin{center}
%\epsfig{file=semall.eps,width=\linewidth}
%\caption{Beam intensity distribution for each year.  The dashed line 
%indicates the boundary between ``low'' and ``high'' intensity for the 
%purposes of our crosscheck.}
%\label{fig:semall}
%\end{center}
%\end{figure}

Figure \ref{fig:epechecks} also shows the value of $\reepoe$ 
for the subsets of data with the regenerator in the left or right
beam.  We find no variation of $\reepoe$ with regenerator position,
which shows that there is no significant intensity difference between
the two neutral beams and that there is no significant left-right asymmetry
in the detector.

There are several crosschecks of the $\kchrg$ sample for which we do not 
divide the $\kneut$ sample.  We divide the $\kchrg$ sample based 
on the polarity of the analysis magnet and whether the tracks bend 
inward or outward
in the magnet.  In each of these cases, the $\kneut$ sample is common to 
both data points and the errors
are estimated by the difference  in quadrature between the subset error 
and the nominal error.  Figure 
\ref{fig:epechecks} shows the $\reepoe$ results for each of these subsets; 
they all show good agreement.
The fit results for tracks that bend in or out in the magnet 
are both slightly larger than the 
nominal result; in this case other fit parameters have changed in each fit 
to allow the higher values of $\reepoe$.

\begin{figure}
\begin{center}
\epsfig{file=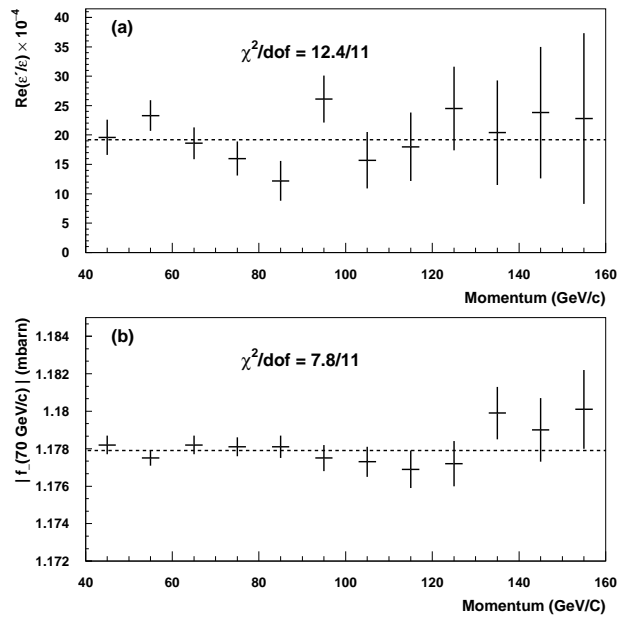,width=\linewidth}
\caption{(a) $\reepoe$ and (b) $\fminus$ in 10 GeV/$c$ momentum bins.
All points are statistically independent.
The dashed lines indicate the values for the full data sample.}
\label{fig:epepbins}
\end{center}
\end{figure}

We check for dependence on kaon momentum by breaking the data into twelve 10 GeV/$c$ momentum bins.  In these
fits, we fix the power-law dependence of the regeneration amplitude to the value found in the nominal fit. The free parameters are $\reepoe$, $\fminus$, and the charged and neutral kaon fluxes.
Figure \ref{fig:epepbins} shows the values of $\reepoe$ and $\fminus$ 
for these fits.  We see
no evidence for dependence of the $\reepoe$ result on kaon momentum.

\begin{figure}
\begin{center}
\epsfig{file=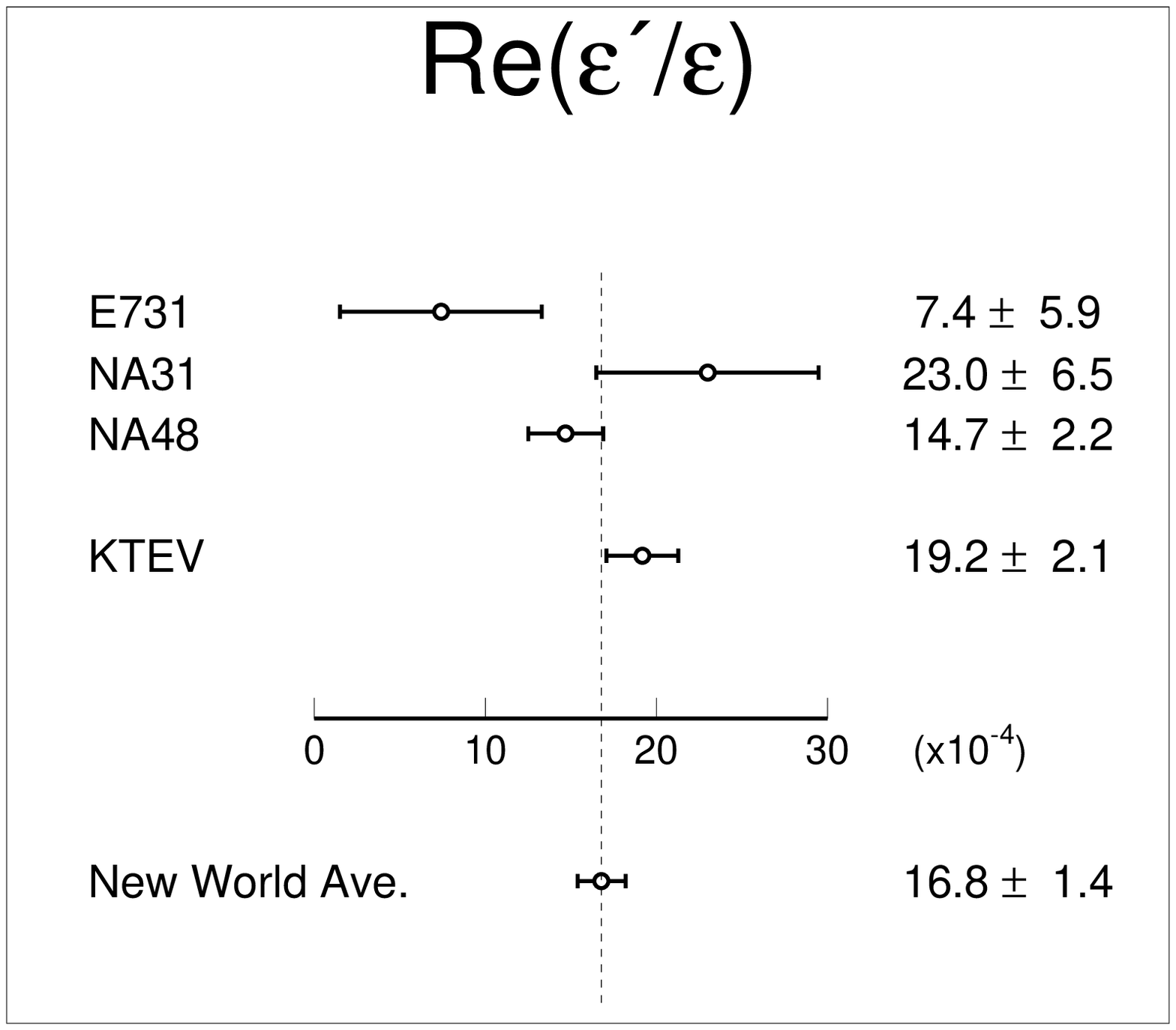,width=\linewidth}
\caption{Final KTeV result for $\reepoe$ and comparisons with previous measurements
from E731\cite{prl:731}, NA31\cite{pl:na31}, and NA48\cite{na48:reepoe}.
The new world average for $\reepoe$ is also shown.}
\label{fig:exptcomp}
\end{center}
\end{figure}

Our value of $\reepoe$ is also consistent with other experimental results
\cite{prl:731,pl:na31,na48:reepoe}.  
The weighted 
average of the new KTeV result with previous measurements is
$\reepoe  =  [16.8 \pm 1.4]\eu$; see Fig. \ref{fig:exptcomp}.  The 
consistency probability of these results is $13\%$.

\subsection{Measurements of Other Kaon Parameters\label{sec:zbinned}}

The regenerator beam decay distribution is sensitive to the
the kaon parameters $\tau_S$, $\delm$, $\phiep$, and
$Im(\epsilon'/\epsilon)$ (see Eq. \ref{eq:reg}).
These parameters can be measured by fitting the decay vertex distribution in 
the regenerator beam. 
The analysis of the kaon parameters follows
the general procedure developed in~\cite{prl:731,prl:pss,prd03}, but uses a
new method to fit for all parameters simultaneously and apply
the CPT constraints {\it a posteriori}.

The  ``$z$-binned'' fit uses both charged and neutral mode data. The data are binned
in $2$ meter wide $z$-bins in the regenerator beam from $124$~m to $158$~m.
A single $z$-bin from $110$~m to $158$~m is used in the vacuum beam.
As in the nominal $Re(\epsilon'/\epsilon)$ fit, there are $24$ free parameters
for the charged and neutral kaon fluxes in each of twelve momentum bins and
two free parameters for the regeneration amplitude and phase.
Two additional parameters are used to fit the effective regenerator edge
for the charged and neutral data to account for smearing effects
near the edge. 

The five kaon parameters, $\tau_S$, $\delm$, $\phiep$, 
$Re(\epsilon'/\epsilon)$, and $Im(\epsilon'/\epsilon)$ are free
parameters of the $z$-binned fit.  The fit thus provides the most
general description of the data with no requirement of CPT invariance.
All systematic uncertainties are evaluated for the fit, accounting 
for correlations among the parameters. CPT invariance is imposed
{\it a posteriori}, including the total errors of the parameters with their
correlations, to obtain precise measurements of $\delm$ and $\tau_S$.

This approach allows a self-consistent analysis of the data with and without
CPT constraints. The results are crosschecked following the procedure developed
in~\cite{prl:731,prl:pss,prd03}, in which  separate fits for $\delm$ and $\tau_S$ were
performed  with CPT invariance  imposed {\it a priori}. 

\subsubsection{Measurement of  $\tauS$, $\delm$, $\phiep$, and $Im(\epsilon'/\epsilon)$ with no CPT constraint}
\begin{figure}
\begin{center}
\epsfig{file=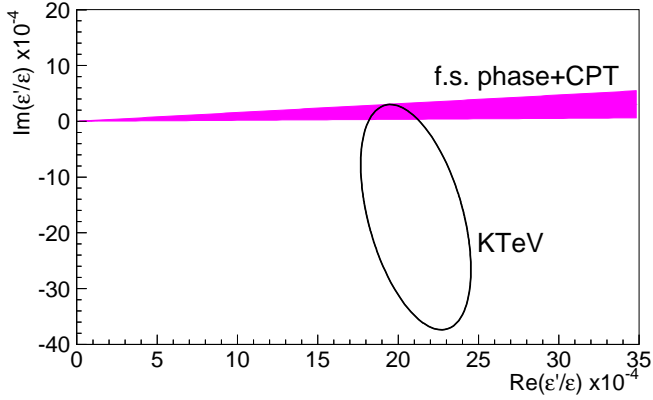,width=\linewidth}
\caption{\label{fig:reim} $\Delta \chi^2 = 1$ contour
for $Re(\epsilon'/\epsilon)$ vs $Im(\epsilon'/\epsilon)$ as measured by KTeV 
compared to the measurement of $\pi\pi$ phase shifts~\cite{ochs} and the CPT 
invariance expectation.
 }
\label{fig:reimcorr}
\end{center}
\end{figure}

The $z$-binned fit results are
\begin{equation} \label{eq:zbinned}
\begin{array}{lcl}
\tauS &=& [\tscptval \pm 0.042_{\rm stat} \pm \tscpterrsy_{\rm syst}] \times 10^{-12}~{\rm s}\\
      &=&  [\tscptval \pm \tscpterr] \times 10^{-12}~{\rm s}\\
\delm &=& [\dmcptval \pm 12.8_{\rm stat} \pm \dmcpterrsy_{\rm syst} ] \times 10^{6}~{\rm \hbar/s} \\
      &=& [\dmcptval \pm \dmcpterr] \times 10^{6}~{\rm \hbar/s} \\
\phiep & = & [\phcptval \pm 0.40_{\rm stat} \pm \phcpterrsy_{\rm syst}]\degs \\
       & = & [\phcptval \pm \phcpterr ]\degs \\
Re(\epsilon'/\epsilon) & = & [\reecptval \pm 1.31_{\rm stat} \pm \reecpterrsy_{\rm syst} ]\times 10^{-4}\\
                       & = & [\reecptval \pm \reecpterr ]\times 10^{-4}\\
Im(\epsilon'/\epsilon) & = & [\imecptval \pm 9.04_{\rm stat} \pm \imecpterrsy_{\rm syst}]\times 10^{-4}, \\
                       & = & [\imecptval \pm \imecpterr ]\times 10^{-4}. \\
\end{array}
\end{equation}
The fit $\chi^2$ is $\chi^2/\nu = 425.4/(432-33)$.
The systematic 
uncertainties are summarized in Table~\ref{tab:zbinsyst}.
The total uncertainty and the correlations among the parameters are 
evaluated following the
procedure described in Appendix D of~\cite{prd03}, and
are given in Table~\ref{tab:corrz}.  The correlations among these
results are shown in Fig.~\ref{fig:swdmts_full} and Fig.~\ref{fig:reim}.

Uncertainties from the charged mode are smaller than those 
from the neutral mode
for the measurements of $\delm$, $\tauS$, and $\phiep$.  The measurement 
of these parameters is effectively a statistical average of the charged 
and neutral mode values, so it is dominated by the statistically larger 
charged mode. The measurements of $\reepoe$ and $\imepoe$ depend on the 
difference between the two modes, so the uncertainties from the 
statistically smaller neutral dataset are more important.  
Similarly, the fitting uncertainties have a much larger impact on $\delm$, 
$\tauS$, and $\phiep$ than on  $\reepoe $ and $\imepoe $ since the 
uncertainties in the regeneration properties enter directly for the former
and cancel for the latter.

There is a large correlation among
$\tau_S$, $\delm$, and $\phiep$, and also between $Re(\epsilon'/\epsilon)$ and
$Im(\epsilon'/\epsilon)$. The correlation between 
$\delm$ and $\phiep$ ($\rho = \dmphcptcor \%$) is
somewhat reduced compared to the pure statistical correlation 
($\rho = 97.3\%$), because the systematic uncertainty due to analyticity
affects $\phiep$ but not $\delm$.
%  primarily because 
% of the analyticity assumption uncertainty which affects $\phiep$ only. 

The measurement of $\imepoe$ can be expressed in terms of the phase difference
\begin{equation}
\begin{array}{lcl}
\Delta \phi \approx  - 3 \imepoe &=& [\dphcptval \pm 0.15_{\rm stat} \pm \dphcpterrsy_{\rm syst}] \degs \\
            &=& [\dphcptval \pm \dphcpterr ]\degs.
\end{array}
\end{equation}
It is consistent with zero as expected from CPT invariance in a
decay amplitude. 
The individual values of $\phipm$ and $\phizz$ are
\begin{equation}
\begin{array}{lcl}
\phipm &=& [\phpmval \pm \phpmerr]\degs\\
\phizz &=& [\phzzval \pm \phzzerr]\degs,
\end{array}
\end{equation}
where the errors correspond to the total uncertainty and are calculated 
including
the correlations reported in Table~\ref{tab:corrz}.

The superweak phase calculated using parameters from Eq.~\ref{eq:zbinned}
is
\begin{equation}
\begin{array}{lcl}
  \phi_{SW} &=&  [\phswcptval \pm 0.069_{\rm stat} \pm  \phswcpterrsy_{\rm syst}]\degs\\
           & =& [\phswcptval \pm \phswcpterr ]\degs.
\end{array}
\end{equation}
The difference
between $\phiep$ and $\phi_{SW}$, 
\begin{equation}
\begin{array}{lcl}
\delta \phi &=& \phiep - \phi_{SW}\\
            &=& [\dphswval \pm   0.37_{\rm stat} \pm  \dphswerrsy_{\rm syst}]\degs \\
            &=& [\dphswval \pm \dphswerr]\degs,
\end{array}
\end{equation}
is also consistent with zero as expected from CPT invariance in $\Kz$-$\Kzbar$
mixing. 
% In this fit,
%the value of $\phi_{SW}$ is computed dynamically using the floated values
% of $\delm$ and $\tauS$.

\begin{table}
\caption{\label{tab:zbinsyst}Systematic uncertainties for the global $z$-binned fit}
\begin{center}
\begin{tabular}{l|ccccc}
\hline
\hline
                                  & $\tauS$ &   $\delm$ &   $\phiep$ & $Re(\epsilon'/\epsilon)$ & $Im(\epsilon'/\epsilon)$ \\
                                  & $\times 10^{-12}$~s &   $\times 10^6~{\rm \hbar/s}$ &   $\degs$ & $\times 10^{-4}$ & $\times 10^{-4}$ \\
\hline  
%%%%  START ZBINSYST
   Trigger                        &   0.004     &     2.4     &      0.08           &        0.13                &      1.16              \\
   Track reconstruction           &            &            &                    &                           &                           \\
   ~~maps                        &  0.000     &   0.0      &   0.00             &      0.04                 &     0.48                  \\
   ~~resolution                  &  0.001     &   2.6      &   0.08             &      0.10                 &     1.20                  \\
   ~~$p_t$ kick                  &  0.009     &   0.7      &   0.00             &      0.14                 &     1.75                  \\
   ~~$Z$ DC                      &  0.002     &   0.1      &   0.00             &      0.28                 &     0.39                  \\
Selection efficiency            &             &            &                    &                           &                           \\
   ~~pt cut                     &   0.008     &   3.6      &    0.10            &    0.16                   &    0.96                   \\
   ~~accidental           &    0.000      &     0.1    &       0.02         &        0.05               &        0.73               \\
   ~~scattering           &    0.001      &     0.3    &       0.10         &        0.15               &        0.17               \\
   Apertures      &            &            &                    &                           &                           \\
~~   Cell separation           &   0.036    &   10.0     &     0.31           &     0.42                  &     2.57                  \\
   Background             &  0.001      &     0.0    &      0.01          &       0.1                 &      0.6                  \\
   Acceptance                                     \\
   ~~Z slope             &   0.007     &   1.4      &        0.04        &        0.13               &     3.05                  \\
\hline
\hline
Trigger                  &   0.002      &   0.9       &       0.02        &          0.08               &       1.71                \\
CsI Reconstruction  \\
~~ Energy linearity     &   0.003       &   0.8       &      0.01         &          2.30              &        2.43                \\
~~ Energy scale         &   0.008       &   0.8       &      0.01        &           1.72              &       12.29                 \\
Selection Efficiency \\
 ~~Ring                   &  0.002      &   0.3       &       0.01        &          0.18               &       2.19                \\
 ~~Pairing $\chi^2$       &  0.012      &  2.2        &       0.07        &          0.02               &       2.19                \\
 ~~Shape   $\chi^2$       & 0.0        &    0.2       &       0.02        &          0.06               &       0.90                \\
Apertures  \\
  ~~CsI size              &  0.006      &    0.2      &       0.04        &         0.64               &        8.35               \\
  ~~MA                     &  0.         &    0.1      &       0.00        &         0.27               &        0.21               \\  
  ~~CA                     &  0.         &    0.2      &       0.01        &         0.47               &        0.32               \\  
 Background             &  0.008      &    0.3      &       0.04        &         0.43               &        6.69               \\    
 Acceptance             &  0.002      &    0.1      &       0.01       &         0.13                &       2.81                \\
\hline
\hline
   Fitting                &             &            &                    &                           &                           \\
   ~~Attenuation Norm     &   0.003     &   0.3      &        0.01        &         0.01              &     0.01                  \\
   ~~Attenuation Slope    &   0.003     &   2.1      &        0.05        &         0.05              &     0.00                  \\
   ~~Target $K_S$         &   0.026     &   4.7      &       0.11         &         0.00              &     0.00                  \\
   ~~Screening            &   0.018     &   5.6      &       0.02         &        0.57               &     1.35                  \\    
   ~~Analytisity          &   0.0       &   0.0      &       0.25         &        0.0                &     0.0                   \\ 
   MC statistics          &   0.016     &   4.9      &        0.15        &        0.36               &     2.78                  \\
\hline
\hline
%%%%  END ZBINSYST
% DAR fit is 116nv9799reimepe_noscreen_amp_darphi
%    ~~Screening            &  +0.056     & -20.7      &      -0.75         &       -0.01               &    +1.44                  \\
   Total Syst             &   \tscpterrsy     & \dmcpterrsy      &  \phcpterrsy    &   \reecpterrsy  &  \imecpterrsy  \\
   Stat Error             &   0.042     &  12.8      &       0.40         &        1.31               &     9.04                  \\
\hline
  Total Error            &    \tscpterr     & \dmcpterr   &  \phcpterr  &        \reecpterr              &    \imecpterr                  \\
\hline

\hline
\end{tabular}
\end{center}
\end{table}

\begin{table}
\caption{\label{tab:corrz}Correlation coefficients for the $z$-binned fit.}
\begin{center}
 \begin{tabular}{lrrrrr}
\hline
\hline
        & $\tauS$ & $\delm$ & $\phiep$ & $Re(\epsilon'/\epsilon)$  \\
\hline
$\delm$                     & $\dmtscptcor\%$   &   \\ 
$\phiep$                    & $\tsphcptcor\%$   & $\dmphcptcor\%$  &     \\
$Re(\epsilon'/\epsilon)$    & $\tsreecptcor\%$  & $\dmreecptcor\%$ &  $\phreecptcor\%$ \\
$Im(\epsilon'/\epsilon)$    & $\tsimecptcor\%$  & $\dmimecptcor\%$ & $\phimecptcor\%$ & $\reeimecptcor\%$ \\
\hline
\hline
\end{tabular}
\end{center}
\end{table}

\subsubsection{Measurement of  $\tau_S$ and $\delm$ with CPT constraint}
\begin{figure}
\begin{center}
\epsfig{file=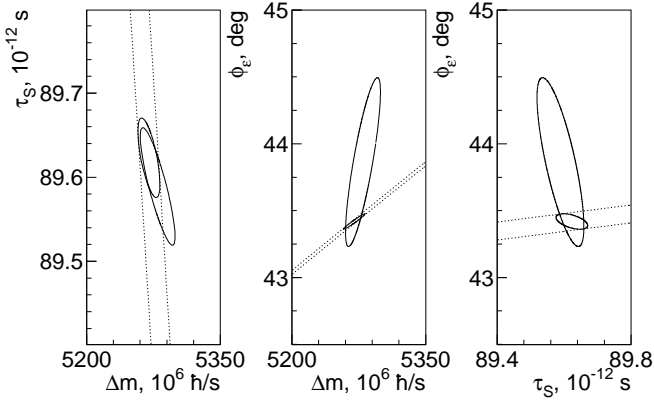,width=\linewidth}
\caption{\label{fig:contours} $\Delta \chi^2 = 1$ contours of total uncertainty
for (a) $\delm$-$\tauS$, (b) $\phiep$-$\delm$ and (c) $\tauS$-$\phiep$.
Larger  ellipses correspond to the $z$-binned fit without  CPT invariance assumption.
Dashed lines correspond to $\phiep = \phi_{SW}$ CPT constraint. Smaller
ellipses are obtained after applying this constraint.
 }
\label{fig:swdmts_full}
\end{center}
\end{figure}

Large correlations of $\delm$ and $\tauS$ with $\phiep$ increase the
experimental uncertainties on these parameters.
Assuming 
\begin{equation}
\begin{array}{c}
\phiep = \phi_{SW}, \\
Im(\epsilon'/\epsilon) = 0,
\end{array}
\end{equation}
as required by CPT invariance,
significantly reduces the errors. This effect is illustrated in 
Fig.~\ref{fig:contours},
which shows $\Delta \chi^2$ contours of total uncertainty for  
$\delm$, $\tauS$, and $\phiep$ 
with and without the CPT constraint.

The results are
\begin{equation}\label{dmts:sw}
\begin{array}{lcl}
  \delm\,|_{\rm cpt}  &=& [\dmswval \pm \dmswerr ] \times 10^6 {\rm \hbar/s} \\
  \tauS\,|_{\rm cpt}  &=& [\tsswval \pm \tsswerr ] \times 10^{-12}~{\rm s}\\
  \rho   &=& \dmtsswcor\%,  \\
\end{array}
\end{equation}
where the errors correspond to the total experimental uncertainty and 
$\rho$ is the correlation coefficient between $\delm$ and $\tauS$.
Compared to the  determination without the CPT constraint, the
uncertainty in $\delm$ is reduced by a factor of $\sim1.5$. Using these
values of $\delm$ and $\tauS$, we determine
\begin{equation}
  \phi_{SW}\,|_{\rm cpt} = [ \phswswval \pm \phswswerr ]\degs.
\end{equation}

\subsubsection{Kaon Parameter Crosschecks}
We compare the new procedure to determine $\delm\,|_{cpt}$ and $\tauS\,|_{cpt}$ 
to the 
one used in~\cite{prd03} in several steps.  First the CPT constraint 
is applied using statistical uncertainties only; these
results agree exactly
with the fits in which the CPT constraints are included as in~\cite{prd03}. 
The value of 
$\delm$ obtained using statistical uncertainties only, 
$\delm\,|_{\rm stat.~cpt} = 5266.5\delmunits$, is somewhat lower than 
for the determination using full uncertainties, Eq.~\ref{dmts:sw}.
This difference can be traced to the above mentioned reduction in the
correlation between $\delm$ and $\phiep$ when systematic uncertainties are 
included.

 Next,  
 $\delm\,|_{cpt}$ and $\tauS\,|_{cpt}$ are determined in the neutral 
and charged modes
separately, as was done in KTeV03\cite{prd03}.  The resulting values  
are $\delm^{00} = (5257.6 \pm 8.3)\times 10^6 {\rm \hbar/s}$, $\delm^{+-} = 
(5269.0 \pm 4.2)\times 10^6 {\rm \hbar/s}$, 
$\tauS^{00} = (89.667\pm 0.039)\times 10^{-12}~{\rm s}$, and 
$\tauS^{+-}= (89.620\pm 0.020)\times 10^{-12}~{\rm s}$, 
where the superscript
$+- (00)$ stands  for the charged (neutral) mode and the errors 
represent
statistical uncertainties only. The measurements agree to within $1.3 
\sigma_{\rm stat}$. 
Finally, the total uncertainties from Eq.~\ref{dmts:sw} are compared to 
an evaluation 
in which the CPT constraints are embedded in the fit. They agree to within 
$\sim 10\%$, which is consistent with small changes in correlations among
the fit parameters.

\subsubsection{Determination of $\Kz$-$\Kzbar$ Mass Difference}
KTeV measurements of the kaon system parameters can be used 
to determine the mass difference between $\Kz$ and $\Kzbar$, which is zero in
the absence of CPT violation. This test uses the Bell-Steinberger relation~\cite{bell-stein},
which connects the CP and CPT violation in the mass matrix to the CP and CPT
violation in the decay. Following the notation used in~\cite{Ambrosino:2006ek}, the 
Bell-Steinberger relation can be written  as
\begin{eqnarray}
\left[\frac{\textstyle \Gamma_S + \Gamma_L} {\textstyle \Gamma_S - \Gamma_L}
+ i \tan \phi_{SW} \right] \left[ \frac{\textstyle Re(\epsilon)}
{\textstyle 1 + |\epsilon|^2 } - i Im(\delta)  \right] = \nonumber \\
= \frac{\textstyle 1}{\textstyle \Gamma_S - \Gamma_L} \sum_f A_L(f)A^*_S(f), \label{eq:bs}
\end{eqnarray}
where the sum runs over all final states $f$ and
 $A_{L,S}(f) \equiv A(K_{L,S} \to f)$. 
The parameter $\delta$ is related to the $\Kz$-$\Kzbar$ mass and decay width difference:
\begin{equation}
\delta = \frac{\textstyle i\left(m_{K^0} - m_{\bar{K}^0}\right)+\frac{1}{2}\left(\Gamma_{K^0}-\Gamma_{\bar{K}^0}\right)}
{\textstyle \Gamma_S - \Gamma_L} \cos \phi_{SW} e^{\textstyle i \phi_{SW}}.
\end{equation}

For neutral kaons, only a few decay modes contribute 
significantly to the sum in Eq.~\ref{eq:bs}. The largest contribution
comes from the $K_{L,S}\to \pi^+\pi^-$ and  $K_{L,S}\to \pi^0\pi^0$
decay modes for which we can use KTeV measurements only; the other required measurements
are the results from Eq.~\ref{eq:zbinned}, and the $K_L\to\pi^+\pi^-$ and 
$K_L\to \pi^0\pi^0$ branching fraction measurements from KTeV~\cite{Alexopoulos:2004sx}. 
The only external input needed is the value of $\tau_L$. We use 
the PDG average, which is based mainly on measurements from KLOE \cite{Ambrosino:2006taul,
Abrosino2005c}.

For the hadronic modes, we define 
\begin{eqnarray}
\alpha_i &\equiv& \frac{\textstyle 1}{\textstyle \Gamma_S}A_L(i)A^*_S(i) = \eta_i B(K_S\to i), \nonumber \\
i &=& \pi^0\pi^0, \pi^+\pi^-(\gamma ), 3\pi^0, \pi^0\pi^+\pi^- (\gamma ).
\end{eqnarray}
For the $K_L\to\pi^+\pi^-$ and $K_L\to \pi^0\pi^0$ decay modes we find
\begin{eqnarray}
  \alpha_{\pi^+\pi^-} &=& \left[ (1124\pm 13) + i (1077 \pm 13 )\right] \times 10^{-6}, \\
  \alpha_{\pi^0\pi^0} &=& \left[ (481\pm 7) + i (465 \pm 7)\right] \times 10^{-6}.
\end{eqnarray}
For the  $K_L\to\pi^+\pi^-\pi^0$ and $K_L\to 3\pi^0$ decay modes we use 
the PDG values
\begin{eqnarray}
  \alpha_{\pi^+\pi^-\pi^0} &=& \left[ (0\pm 2 ) + i (0 \pm 2)\right] \times 10^{-6}, \\
  |\alpha_{\pi^0\pi^0\pi^0}| &<& 7\times 10^{-6}~\mbox{at~}95\%~\mbox{C.L.}.
\end{eqnarray}
For the semileptonic decay modes we use the definition and values from
the PDG
\begin{eqnarray}
\alpha_{\pi l \nu} \equiv \frac{\textstyle 1}{\textstyle \Gamma_S} \sum_{\pi l \nu} A_L(\pi l \nu)A^*_S(\pi l \nu) + \nonumber \\
+ 2i \frac{\textstyle \Gamma_L}{\textstyle \Gamma_S}B(K_L\to \pi l \nu)Im(\delta)
\end{eqnarray}
and
\begin{equation}
\alpha_{\pi l \nu} = \left[ (-2 \pm 5) + i(1 \pm 5) \right] \times 10^{-6}.
\end{equation}
Combining the experimental data, we find from Eq.~\ref{eq:bs} 
\begin{equation}
 Im(\delta) = ( \dmkzkbimv \pm \dmkzkbime ) \times 10^{-5}. \label{eq:imd}
\end{equation}
The uncertainty in $Im(\delta)$ is $1.9\times 10^{-5}$ 
in the PDG average; the reduction of this error in Eq.~\ref{eq:imd}
comes mainly from
the more precise measurement 
of $\phi_{\epsilon}-\phi_{SW}$ from KTeV.

\begin{figure}
\centerline{\epsfig{file=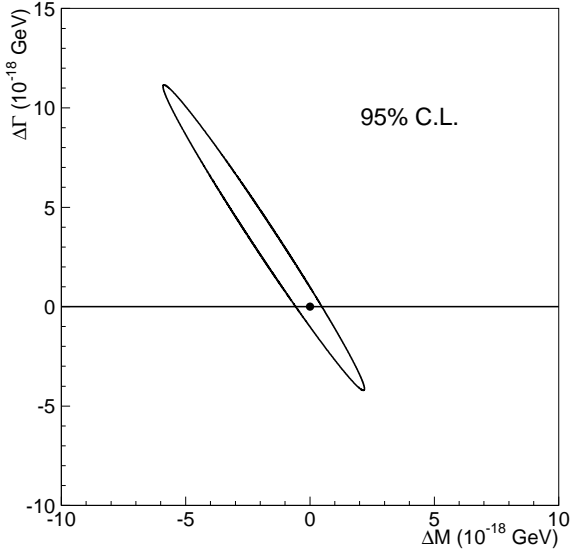,width=0.9\linewidth}}
\caption{Allowed region at $95\%$ C.L. for $\Delta M$, $\Delta \Gamma$. \label{fig:dmdg}}
\end{figure}
Combining the new determination of $Im(\delta)$ with the determination of 
$Re(\delta)$ and $Im(\delta)$ from the charge asymmetry in semileptonic 
decays~\cite{pdg},
\begin{eqnarray}
 Re(\delta) &=& (30\pm 23)\times 10^{-5}, \nonumber \\
 Im(\delta) &=& (-660\pm 650)\times 10^{-5}, \nonumber\\
 \rho &=& -21\%,
\end{eqnarray}
we find 
\begin{eqnarray}
 Re(\delta) &=& ( \dmkzkbfrefv \pm \dmkzkbfrefe )\times 10^{-5}, \nonumber \\
 Im(\delta) &=& (\dmkzkbimfv  \pm  \dmkzkbimfe )\times 10^{-5}  \label{eq:deltaa}
\end{eqnarray}
with negligible correlation between the real and imaginary parts.

Using the value of $\delta$ from Eq.~\ref{eq:deltaa}, we derive the allowed region for $\Delta M \equiv m_{K^0} - m_{\bar{K}^0}$ and 
$\Delta\Gamma \equiv \Gamma_{K^0} - \Gamma_{\bar{K}^0}$ as shown in Fig.~\ref{fig:dmdg}.
Assuming no CPT violation is present in the decay amplitudes, i.e. $\Delta\Gamma=0$, we obtain
the limit for the mass difference,
\begin{equation}
| M_{K^0} - M_{\bar{K}^0} | <  \dmkzkblimit \times 10^{-19}~\mbox{GeV/$c^2$ at } 95\%~\mbox{C.L.}
\end{equation}

\section{\label{sect:conclude}Conclusions}
Using the full data sample of the KTeV experiment, 
we have made improved
measurements of the direct CP violation parameter, $\reepoe$,
and other parameters
of the neutral kaon system.
All of these results supersede previous KTeV results.

Assuming CPT invariance, we measure the direct CP violation parameter
\bqa
\reepoe  & = & [19.2 \pm 1.1(stat) \pm 1.8(syst)]\eu \\ \nonumber
        & = & [19.2 \pm 2.1]\eu.  
\eqa
Also under the assumption of CPT invariance, we report new measurements
of the $\KL-\KS$ mass difference and the $\KS$ lifetime:

\begin{equation}
\begin{array}{lcl}
  \delm  &=& [\dmswval \pm \dmswerr ] \times 10^6 {\rm \hbar/s}\\
  \tauS   &=& [\tsswval \pm \tsswerr ]\times 10^{-12}~{\rm s} . \\
\end{array}
\end{equation}

To test CPT symmetry, we measure the phase difference between the
$\kchrg$ and $\kneut$ decays to be
\begin{equation}
\begin{array}{lcl}
\Delta \phi &=& - 3 \imepoe\\
%            &=& [\dphcptval \pm 0.15_{\rm stat} \pm \dphcpterrsy_{\rm syst}] \degs \\
            &=& [\dphcptval \pm \dphcpterr ]\degs,
\end{array}
\end{equation}
and the phase difference relative to the superweak phase to be
\begin{equation}
\begin{array}{lcl}
\phiep - \phi_{SW}            
%&=& [\dphswval \pm   0.37_{\rm stat} \pm  \dphswerrsy_{\rm syst}]\degs \\
            &=& [\dphswval \pm \dphswerr]\degs.
\end{array}
\end{equation}
These phase results are consistent with CPT invariance in both the decay 
amplitudes and $\Kz-\Kzbar$ mixing.  Assuming no CPT violation in the decay 
amplitudes, we set a limit on the $\Kz-\Kzbar$ mass difference:
\begin{equation}
| M_{K^0} - M_{\bar{K}^0} | <  \dmkzkblimit \times 10^{-19}~\mbox{GeV/$c^2$ at } 95\%~\mbox{C.L.}
\end{equation}

After decades of experimental effort, direct CP violation in the neutral kaon
system has now been measured with an uncertainty of about 10\%. 
%A weighted
%average of
%the new KTeV result with previous measurements gives
%$\reepoe  =  [16.8 \pm 1.4]\eu$ with a 13\%\ confidence level. 
%It is interesting to note that this result corresponds to a
%particle-antiparticle partial decay rate asymmetry of
%\begin{equation}
%\frac{\Gamma(\Kz \to \pi^+\pi^-) - \Gamma(\Kzbar \to \pi^+\pi^-) }
%{\Gamma(\Kz \to \pi^+\pi^-) - \Gamma(\Kzbar \to \pi^+\pi^-)}
%= (5.4 \pm 0.5) \times 10^{-6}. 
%\end{equation}
%for $K \to \pi^0\pi^0$, the corresponding asymmetry is $-2$ times this result.
%The comparable direct CP violation rate asymmetry measured in $B$ decay is
%about $10\%$.
Considerable improvement in theoretical calculations of $\reepoe$
will be required to take advantage of this experimental precision. 
There is some optimism~\cite{Buras:2003zz}, however, that future calculations 
using lattice gauge theory may approach a 10\%\ uncertainty, making the 
precise measurement of $\epe$ an equally precise test of the Standard Model.

\begin{acknowledgments}
We gratefully acknowledge the support and effort of the Fermilab
staff and the technical staffs of the participating institutions for
their vital contributions.  This work was supported in part by the U.S.
Department of Energy, The National Science Foundation, The Ministry of
Education and Science of Japan,
Funda\c c\~ao de Amparo a Pesquisa do Estado de S\~ao Paulo-FAPESP,
Conselho Nacional de Desenvolvimento Cientifico e Tecnologico-CNPq and
CAPES-Ministerio Educa\c c\~ao.
\end{acknowledgments}

\appendix

\section{Screening Corrections to Kaon Regeneration and Analyticity \label{sec:regeneration}}
The measurements of $\phiep$ and $\delm$ rely on an accurate determination
of the regeneration phase. The regeneration phase is calculated from the momentum
dependence of the regeneration amplitude using an analyticity relation.
Regeneration in carbon, the principal material of the KTeV
regenerator~\footnote{Total regeneration amplitude of the KTeV regenerator
includes small corrections from regeneration in hydrogen and lead.
For simplicity, these corrections are neglected in the following discussion.}, 
is discussed here.
We show how the momentum dependence of the regeneration amplitude,
which is calculated based on Regge theory with nuclear screening corrections,
is checked using $K\to\pi\pi$ data. Based on this study, we derive 
 the systematic uncertainties  of the kaon parameters which are correlated
with the determination of the regeneration phase.

In Regge theory, for an isoscalar target and momenta above $10$~GeV/$c$,
the momentum dependence for the magnitude 
of  the forward kaon-nucleon scattering amplitude 
difference  for $\Kz$ and $\Kzbar$
 (Eq.~\ref{eq:fminus}) is given by a single
power law~\cite{pr:gilman}:
\begin{equation} \label{eq:power}
|f^P_-(p)| = |f^P_-(70~{\rm GeV/}c)|\left(\frac{\textstyle p}{\textstyle 70\, \
{\rm GeV/}c}\right)^{\alpha}.
\end{equation}
Analyticity of the forward scattering amplitudes leads to a kaon momentum independent phase of $f^P_-$ given by
\begin{equation} \label{eq:analphi}
\arg(f^P_-) = -\pi\left(1 + \frac{\textstyle \alpha}{\textstyle 2}\right).
\end{equation}
Therefore, $\alpha$ and $|f^P_-(70~{\rm GeV/}c)|$ determine $f^P_-(p)$ fully;
they 
enter  as free parameters in fits to KTeV data.

\begin{figure}
\begin{center}
\epsfig{file=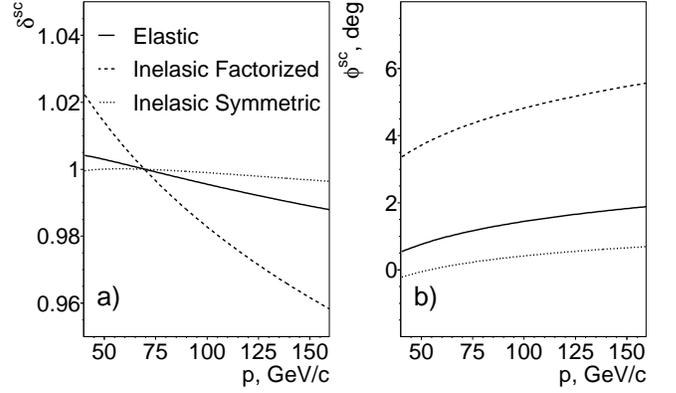,width=\linewidth}
\end{center}
\caption{\label{fig:screen}
Kaon momentum dependence of screening corrections to (a:) the magnitude  and  (b:) the  phase of $f_-(p)$, for various models from \cite{regmodels}. The magnitude of each correction is defined to be 1 at 70 GeV/$c$.}
\end{figure}

Kaon-nucleon interactions in carbon are screened
due to rescattering processes. 
The effect of screening modifies the momentum dependence of 
$|f_{-}(p)|$ as well as its phase:
%The screening corrections are applied to the amplitude and phase of $f_-$ as a multplicative and additive
% momentum dependent correction:
%The total $f_-$ is calculated as a product of a pure power
%law and the momentum dependent screening correction 
\begin{equation} \label{eq:screen}
%\begin{array}{lcl}
f_-(p) =f^P_-(p) \cdot \delta^{\rm sc}(p) \,  e^{\textstyle i \phi^{\rm sc}(p) }.
%\end{array}
\end{equation}
Here $\delta^{\rm sc}(p)$ and $\phi^{\rm sc}(p)$ stand for 
screening corrections, which are evaluated using Glauber theory 
formalism~\cite{glauber55,glauber66} 
for diffractive scattering,
and using various models~\cite{regmodels} for inelastic scattering.
% The regeneration in hydrogen causes a small $<0.1\degs$  
%\begin{equation}
%\begin{array}{lcl}
%   |f_-(p)|    &=&  |f^P_-(p)| \times \delta^{sc}(p)  \\
% \arg(f_-(p))  &=& \arg(f^P_-(p)) + \phi^{sc}(p), 
%\end{array}
%\end{equation}
% are shown in Fig.~\ref{fig:screen}. 
Elastic screening corrections are calculated with small theoretical 
uncertainties.
Inelastic corrections, however, have a large spread of predictions as shown
 in Fig.~\ref{fig:screen}.
Since a momentum independent change of $|f_-|$
does not modify $\arg(f_-)$, calculations of the screening corrections fix
$\delta^{\rm sc}(p)=1$  for $p=70$~GeV/$c$.
Screening corrections calculated using the inelastic factorized 
model~\cite{regmodels} 
are used for central values of all KTeV results in this paper.
% and refer to this  model simply as
%inelastic screening unless explicitly specified.

% and estimate 
%the systematic uncertainty due to smearing using inelastic-symmetric (Inelastic Sym) model
%Regeneration in  plastic scintillator, in the following reffered to as 
%$f^{\rm scint}_-$, acquires  
% a small additive correction from regeneration in hydrogen.

\begin{figure}
\begin{center}
\epsfig{file=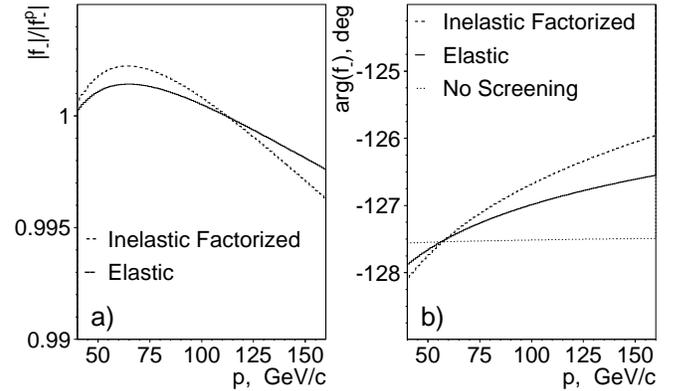,width=\linewidth}
\end{center}
\caption{\label{fig:screenfit}Kaon momentum dependence of
(a:) ratio of $|f_-(p)|$
from the fits using elastic (solid line) and inelastic factorized 
(dashed line) screening 
corrections to $|f^P_-(p)|$ from the fit using no screening corrections;
(b:) $\arg(f_-(p))$ from the fits with elastic (solid line), inelastic (dashed line)
screening corrections and with no screening corrections (dotted line). 
}
\end{figure}
%Note that some of the differences in the corrections 
%Significant differences 
As shown in Fig.~\ref{fig:screen}, screening corrections
modify the value of effective power law $\alpha$ from Eq.~\ref{eq:power}. 
Since $\alpha$ is a free parameter of  fits to the data,
it is adjusted such that the net effect of the screening corrections is reduced.
Figure~\ref{fig:screenfit}a shows ratios of fitted~\footnote{All fits discussed
in this  section are $z$-binned fits with no CPT constraint,
see Sec~\ref{sec:zbinned}.}  
$|f_-(p)|$ using elastic and inelastic factorized screening models
to fitted $|f^P_-(p)|$ using no screening corrections.
Figure~\ref{fig:screenfit}b shows the
phase, $\arg(f_-(p))$, obtained in these fits~\footnote{Note that small variation of the phase for the fit using no screening corrections
 is caused by the regeneration in hydrogen.}.
The screening correction thus results in
up to a $0.5\%$ adjustment of the regeneration amplitude and up to a $2\degs$ change in the 
regeneration phase, for the momentum range of the analysis. 
The values of $\alpha$ obtained in the fits with
different screening models are given in Table~\ref{tab:alpha}. Comparing
$\chi^2/\nu$ for these fits, one can see that the data rule out calculations
without screening corrections and disfavor the inelastic symmetric screening
 calculation.

\begin{table}
\caption{Power law coefficient $\alpha$ and $\chi^2/\nu$ for fits using
various screening models\label{tab:alpha}. The uncertainties
given in parenthesis are statistical errors.}
\begin{center}
\begin{tabular}{l|cc}
\hline
\hline
                         &  $\alpha$           &  $\chi^2/\nu$ \\
\hline
No screening             & $-0.5813(5)$        & $471/399$  \\
Elastic screening         &  $-0.5715(5)$      & $431/399$ \\
Inelastic factorized screening  & $-0.5376(5)$ & $425/399$  \\
Inelastic symmetric screening  & $-0.5803(5)$  & $438/399$  \\
\hline
\hline
\end{tabular}
\end{center}
\end{table}

Comparison of Fig.~\ref{fig:screen} and Fig.~\ref{fig:screenfit} suggests
significant sensitivity to the regeneration parameters and that the
screening corrections may be determined directly from a fit to the $K \to \pi\pi$
 data. 
%Here we improve the error estimation 
%using $K \to \pi\pi$ data to verify the screening calculations.
% The caclulation of the screening corrections 
% is checked using $K_L \to \pi\pi$ data.
In this fit, the magnitude of the regeneration amplitude,
$|f_-(p)|$, is floated as a free parameter in≈ß each of the twelve
 $10$~GeV/$c$ momentum bins, and
is approximated to be constant within each bin.
The regeneration phase,
$\arg(f_-(p))$, is calculated dynamically
using the Derivative Analyticity Relation (DAR)~\cite{PhysRevD.12.3431}:
\begin{equation} \label{eq:localanal}
 \arg(f_-(p)) = -\pi - \tan(\frac{\pi}{2} \frac{{\rm d}} {{\rm d} \ln p}) 
\ln |f_-(p)|.
\end{equation}
In this calculation, $\arg(f_-(p))$ is estimated  
at an average of bin centers for the neighboring momentum bins,
assuming that the bin-to-bin dependence  
of $|f_-(p)|$ is described by a 
power law.
For momenta away from the bin centers,
the phase is interpolated linearly. 
For example, the local power law at $p=50~{\rm GeV/}c$ is calculated as
\begin{equation}
   \alpha (50~{\rm GeV/}c) = -  
\frac{\textstyle 
\ln \frac{\textstyle |f_-(55~{\rm GeV/}c)|} {\textstyle |f_-(45~{\rm GeV/}c)|}}
{\textstyle \ln  \frac{\textstyle 55 } {\textstyle 45} }.
\end{equation}
The phase $\arg(f_-(50~{\rm GeV}))$ is then given by
Eq.~\ref{eq:analphi}.
%The other parameters of the fit are the same as in $z$-binned fit described in
%Sec.~\ref{sec:zbinned};
This fit is referred to as the ``DAR fit''. Compared to the standard $z$-binned
fit, the DAR fit has an additional $12-1$ free parameters.

\begin{figure}
\begin{center}
\epsfig{file=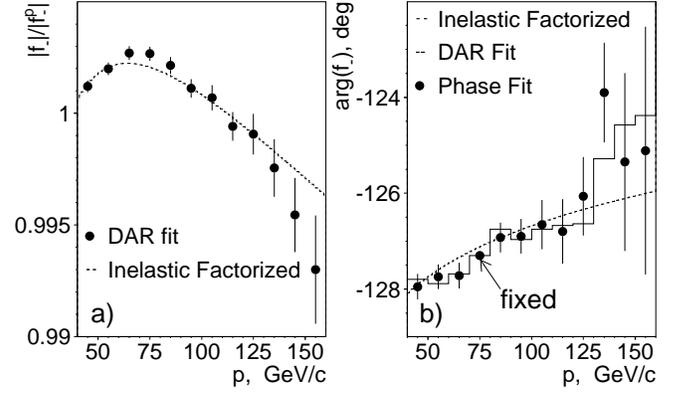,width=\linewidth}
\end{center}
\caption{\label{fig:dar}
Kaon momentum dependence of (a:) ratio of $|f_-(p)|$ 
from  the fit
 using inelastic factorized screening corrections (dashed line)
 and from the DAR fit (dots with error bars) to
$|f^P_-(p)|$ from 
 the fit using no screening correction; (b:) $\arg(f_-(p))$ 
for the DAR fit (solid line), the phase fit (dots with error bars)
and the fit using inelastic factorized screening model (dashed line). 
The phase fit is fixed to the DAR value for $70$~GeV/$c<p<80$~GeV/$c$ 
momentum bin as indicated by the arrow.}
\end{figure}
Figure~\ref{fig:dar}a compares ratios of $|f_-(p)|$ obtained from the DAR fit
and from the fit using the inelastic factorized screening correction to $|f^P_-(p)|$ obtained
from
the fit using no screening corrections. There is a good agreement between
the inelastic screening calculation and the DAR fit.
%The ratio $|f_-|(p)/|f_-^P|(p)$ determined
%from the DAR fit agrees well with the inelastic screening calculation
%as can be seen in Fig.~\ref{fig:dar}a.
%A comparison of the calculated screening corrections and the data
%are shown in Fig.~\ref{fig:dar}. The left (right) panel compares the 
% correction to $|f_-|$ ($\arg(f_-)$).  
%The absolute
%value of the forward scattering amplitude difference  
%agrees well between the DAR fit and inelastic screening
The quality of the DAR fit   ($\chi^2/\nu = 412/388$)  
is similar to that of the fit using inelastic screening corrections.

The data precision makes it possible to go one step further and float both
$|f_-(p)|$ and $\arg(f_-(p))$ for each momentum bin, this fit is called
the ``phase fit''.
%The quality of the DAR prediction for $\arg(f_-)$
% is tested  using a  $z$-binned fit in which, in addition
%to $|f_-(p)|$, $\arg(f_-(p))$ are 
%% also float as free parameters for each momentum bin. 
%This fit is referred to as the ``phase fit''.
An overall phase  can not be measured this way, 
since it is $100\%$ correlated with the value
of  $\phi_{\epsilon}$. Therefore
 the phase is fixed at $p = 75$~GeV/$c$  to the value obtained in
 the DAR fit.
The phase fit thus does not check an absolute prediction of the  DAR fit
but only the  
predicted momentum dependence. The quality of the phase fit is not improved
compared to the DAR fit: for extra $11$ degrees of freedom $\chi^2$ is 
reduced by  $7$ units only.
Figure~\ref{fig:dar}b compares $\arg(f_-(p))$  calculated in the DAR fit and in the fit using the factorized 
inelastic screening correction
as well as  directly fitted in the phase fit.
These three different determinations are consistent with each other.

Note that the large correlation between   $\arg(f_-)$ extracted
at different momenta in the phase fit complicates a quantitative comparison
between the predictions, 
so an
additional test has been performed.
The  variation of $\arg(f_-)$ versus momentum, 
predicted by the DAR fit, shows an approximately linear dependence. The total
variation, for the $40-160$~GeV/$c$  momentum range, is about $3\degs$.
As a crosscheck, linear dependence of $\arg(f_-)$ is assumed: 
\begin{equation}
\arg(f_-(p)) = \arg(f_-^P) + \alpha_R \cdot \frac{p - 70~{\rm GeV/}c}{120~{\rm GeV/}c},\\
\end{equation}
 where $p$ is 
in GeV/$c$ and $\alpha_R$ is an additional free fit parameter.  
A fit with $|f_-(p)|$  floated in twelve 
momentum bins using no screening corrections 
leads to significantly 
non-zero $\alpha_R$: $\alpha_R = (3.1 \pm 0.6)\degs$. A similar  fit, which 
includes the
DAR phase correction, gives $\alpha_R$ consistent with zero: $\alpha_R = (0.2 \pm 0.6)\degs$.

In KTeV03~\cite{prd03}, the systematic uncertainties
due to the screening corrections were evaluated by comparing the factorized and symmetric inelastic  screening models
which led to a large $0.75\degs$ error for $\phiep$. 
In this analysis, the systematic uncertainties 
are determined directly from the data and 
estimated
by comparing the fit based on the inelastic factorized screening calculation with the DAR fit.
For $\phiep$, this evaluation leads to a small uncertainty 
of $ 0.02\degs$, see Table~\ref{tab:zbinsyst}.
%,  since the overall correction to  the
%regeneration phase agrees well for   $p<130$~GeV/$c$  where the bulk of the data
%statistics lie. 
The uncertainty is larger
for $\delm$ ($5.6\times10^{-12}~{\rm \hbar/s}$), since $\delm$ is more sensitive
to the variation of $\arg(f_-)$ as a function of the kaon momentum.

The determination  of the  regeneration phase relies on the validity of 
the analyticity assumption,
Eq.~\ref{eq:analphi} 
(or equivalently of Eq.~\ref{eq:localanal}).
%The systematic uncertainty from the analyticity assumption for the forward
%scattering amplitude is taken fro
Various sources of deviation from Eq.~\ref{eq:analphi} 
were studied in~\cite{analphi}. The net effect
of sub-leading Regge trajectories, uncertainties in  $f_-$ 
at low momentum, and electromagnetic regeneration in lead were estimated to be
below $0.25\degs$, which is taken as a systematic uncertainty for the 
regeneration phase. Note that an exchange of an Odderon can affect
the  regeneration amplitude at high momentum, but  is not taken into account in this evaluation. 
The Odderon contribution can be limited using $pp$ and $p\bar{p}$ data; the best
fit to these data gives a regeneration phase shift of $-(0.2\pm 0.6)\degs$~\cite{analphi}.  

A cross check of the analyticity assumption is performed by measuring the asymmetry between
$K\to \pi^-e^+\bar{\nu}_e$ and $K\to \pi^+e^-\nu_e$ decays downstream of the regenerator
which is sensitive to the regeneration phase.
Using approximately  $125$ million decays, the difference between the measured
$\arg(f_-)$ and the prediction based on  analyticity is 
$[-0.70\pm0.88 (stat) \pm 0.91 (syst)]\degs$~\cite{thesis:gcb},
which is consistent with the analyticity assumption.

\bibliography{prd08_bib}% Produces the bibliography via BibTeX.

%\end{linenumbers}
\end{document}